\documentclass{elsart}
\usepackage{epsfig}
\begin{document}

\begin{frontmatter}
\title{Deep Level Promotion Mechanism in Sputtering}
\author{A. Sindona\thanksref{e-m1}} 
and
\author{G. Falcone\thanksref{e-m2}}
\address{
         Dipartimento di Fisica, 
         Universit\`a della Calabria,
         Arcavacata di Rende, 
         87030 Cosenza, Italy
         }

and

\address{
         Unit\`a dell'Istituto Nazionale di Fisica          
         della Materia(INFM) di Cosenza,
         Italy
         }
\thanks[e-m1]{e-mail: Sindona@fis.unical.it}
\thanks[e-m2]{e-mail: Falcone@fis.unical.it}
\begin{abstract}
We have applied a double decoupled localized level Anderson-Newns Hamiltonian to the analysis of surface effects upon the ionized fraction $\mathcal{R}_{+}$ of sputtered atoms from a metal surface. 
Electronic excitations, induced in the conduction band by the transient formation of quasi molecular systems, between substrate and emitted atoms, in the collision cascade generated by the primary incident beam, have been explicitly included into an instantaneous transition matrix peaked at the Fermi level of the material. 
The interaction dynamics seem to take place over two different time scales, one related to sputtered atom trajectories and the other to recoiled substrate particles. 
Finite temperature calculations have suggested, at very low ejection energies, a power law dependence of the final charge state of the sputtered beam on its detected velocity. 
This result is in agreement, in the zero temperature limit, with some previously published papers and its validity has been compared to other theoretical outcomes and tested on SIMS data.

\textit{pacs}: 79.20.-m, 71.23.An

\textit{keywords}: ion-solid interactions; 
                   ion emission;                    
                   metallic surfaces; 
                   models of non equilibrium phenomena;                                       scattering and sputtering.
\end{abstract}
\end{frontmatter}

\section{Introduction}

Resonant electron transfer during sputtering of atoms and molecules from metal surfaces has been a very interesting subject in relation to secondary ion mass spectroscopy (SIMS) where the knowledge of ionization and neutralization rates is essential for many quantitative analyses. 
The theory of such processes should encompass a detailed description of charge exchanges in the collision cascade to the outgoing atom or atomic cluster.
In fact, to make this complicated problem tractable, a basic model has been formulated in terms of a single orbital parametrically time-dependent Anderson-Newns 
Hamiltonian {\cite{Anderson}-\cite{Falcone}}, in which the effects of the bombardment, including the substrate motion, the electronic excitations generated and the physical and chemical consequences of the void produced by the ejected particle have been neglected. 
Many experimental results have been reproduced in this way {\cite{Yu}-\cite{Overbosh}} which is
common to both SIMS and atomic surface scattering. 
Yet the need for a more comprehensive theory of surface influences in sputtering has led to a generalized approach {\cite{Falcone,Sroubek3,Sroubek4,Anto}} in which the substrate dynamics have been modeled with locally time-dependent perturbations created during the collision and acting within the first layers of the surface region. 
In a first simplified treatment \cite{Sroubek3} a one-body time dependent scattering potential, localized in the vicinity of the emission site, was added to the basic Hamiltonian and \textit{direct} (i.e. non mediated by the atomic localized level) electronic excitations, created in the conduction band by this new term, were explicitly considered in the evaluation of the ionized fraction. 
Although the average lifetime of these excitations is believed to be rather short on the time-scale of the emission trajectory  \cite{Lang3}, their contribution to the final charge state of secondary ions becomes important at very low emission energies where the ionization probability calculated from the basic theory exhibits an exponential dependence on the inverse of the \textit{average} outgoing velocity of sputtered particles {\cite{Blandin,Brako1}}. 
In a recent paper \cite{Anto}, an exact formal solution of the two potential problem formulated in Ref.\cite{Sroubek3} has been obtained confirming more rigorously that, in the velocity regime typically observed in many SIMS experiments, the basic hopping mechanism may be negligible with respect to surface induced excitations, at least if they can be represented in the form of a one-body scattering potential. 
In the present article, we have proposed a more accurate form of the sputtering potential combining the above mentioned impurity scattering contribution with a hopping resonant interaction due to a second discrete state embedded in the conduction band (Sec.2). 
Both perturbations are essentially generated in the quasi-molecules (QM) formed \textit{transiently} in the collision cascade, between substrate and emitted atoms, and the
latter has been already used in a different context concerning the analysis of electronic emission from a bombarded metal substrate \cite{Falcone,Sroubek4}. 
The reference Hamiltonian is quadratic in electron creation and annihilation operators, which would make the problem trivially soluble in the absence of time-dependent interactions. 
Nevertheless, in the present treatment, only a formal expression can be given to the exact evolution of the localized orbital on the ejected particle. 
We have performed this step by using retarded Green's functions thus parameterizing the evolution of the atomic state in terms of Feynmann diagrams (Sec.3). 
Assumptions of the slow, non adiabatic dynamics, i.e. Brako and Newns' approach \cite{Brako1,Brako2,Brako3}, plus an approximation due to the transient nature of the QM, has allowed us to obtain a manageable expression for multiple interaction transition amplitudes involved in the problem (Sec.4). 
We have generalized former results \cite{Sroubek3,Anto}, with particular attention to the dependence of the ionization probability on the final emission velocity of the outgoing
ion, including thermal interactions. 
We have also proposed a comparison of our predictions with previously published theoretical calculations
  \cite{Sroubek1,Blandin,Brako1,Sroubek2,Sroubek3,Veksler} and applied them to a well known sputtering experiment   {\cite{Sroubek2,Hart}}.

\section{Formulation of the Problem and Generalized Hamiltonian.}

The generalized model is sketched in Fig.1. The metal is treated as a Fermi gas specified by a set of continuous states $\left\{ \left| k\right\rangle \right\}$ and spectrum 
$\left\{ \varepsilon _{k}\right\}$ with work function $\phi$. 
The Fermi energy has been set to $\varepsilon_{f}=-\phi$ in an appropriate scale with the \textit{vacuum} fixed to zero. 
The multiple index $k$ runs over single particle energies and all the degeneracies of
each level. 
In addition, because of impurities, adatoms and surface defects, it may include discrete quantum numbers. 
The outward atom is assumed to move along a classical trajectory and have a non degenerate valence level 
$\varepsilon _{a}\left( t\right)$ lying within the metal conduction band.
The corresponding valence state is indicated by $\left|a\left(t\right)\right\rangle$. 
$\varepsilon _{a}$ may be an \textit{affinity level} ($A$) as well as an \textit{ionization level} ($I$) which give rise, respectively, to negative and positive ionization. 
An electron can tunnel between the atomic level and a continuous state 
$\left| k\right\rangle $ with a  coupling strength given by the hopping integrals {\cite{Anderson,Newns}} $V_{ka}\left(t\right)$ and 
$V_{ak}\left( t\right) = V_{ka}^{\ast }\left( t\right)$.
Particle exchange with the continuum dominates in a range of the order of the virtual \textit{broadening} of the localized state 
\begin{equation}
\Delta_{a}\left( t\right) = 
\pi \sum\limits_{k}
\left|V_{ka}\left(t\right)\right|^{2}
\delta \left[ 
\varepsilon _{k}- \varepsilon _{a}\left( t\right) 
\right] ,  
\label{Eq:2.1}
\end{equation}
which is proportional to the first order transition rate of all occupation channels 
$\left| k\right\rangle \rightarrow \left| a\right\rangle $, i.e Fermi's golden rule \cite{Gadzuk}. 
Anderson-Newns' potential causes an \textit{indirect} production of electron-hole pairs by processes \cite {Brako2,Falcone,Sroubek3,Sroubek4,Muller} of the kind shown in Fig.2a. 
Other excitation channels are induced by \textit{transient perturbations} generated during the collision process. 
In relation with the schematic of Fig.3, we observe that the majority of low energy atoms are emitted because of \textit{indirect binary collisions} with substrate atoms that have received a sudden impulse in the collision cascade and the number of ejections caused by \textit{direct collision} with primary ions may be neglected.
For this reason, we consider secondary emitted atoms to form quasi molecular systems with recoiling substrate atoms moving within the surface region. 
With this mechanism an electron may be \textit{trapped}, as illustrated in Fig.4, into a transient \textit{molecular orbital} (MO) which may be promoted to the conduction band as a result of the increasing nuclear charge in the QM with decreasing internuclear distance and the increase in kinetic energy of the electron with its increased localization. 
Thus the hopping interaction of this state with the continuum can be very important at low incident energies when the QM has a longer average lifetime. 
In addition, since translational invariance is broken down both by the substrate and secondary atomic motion, many scattering transitions occur in the conduction band together with electronic emissions.

In this paper we have limited our investigation to the influence of a single promoted state 
$\left| m\left( t\right) \right\rangle $, with energy  $\varepsilon _{m}\left( t\right)$ in resonance with the conduction band, interacting through a second hopping potential, specified by matrix elements  
$V_{km}\left( t\right)$ and 
$V_{mk}\left( t\right) =V_{km}^{\ast }\left(t\right) $, with the continuum. 
This MO may be either a metastable configuration of the outgoing atomic state or distinct from the atomic state itself, depending on the energy position of the localized levels of the two atoms involved in its formation.

Let us denote by $\vec{X}\left( t\right)$ and 
$\vec{Y}\left( t\right)$ the classical position vectors of the substrate and emitted atoms in the QM, respectively, as measured from a reference frame placed at the surface.
Because of the slow motion of the atomic cores on the electronic time-scale, $\vec{X}$ and $\vec{Y}$ introduce two large characteristic times, $\tau _{X}$ and $\tau _{Y}$, over which many electronic transitions occur. 
Moreover, since the substrate recoling atom in the QM, see Fig.3, is more energetic than the emitted atom and follows a complicated path inside the metal, we may take  
$\tau _{Y}\gg \tau _{X}$.

In order to construct the second quantization Hamiltonian of the problem we consider a single electron of the system, with kinetic energy $\mathrm{e}_{K}=\mathrm{p}^{2}/{2}$, $\vec{\mathrm{p}}$ being its momentum operator,  in the surface barrier of the metal 
$v_{s}\left( \mathrm{z}\right)$,  with $\mathrm{z}$ its orthogonal coordinate operator, and in the screened Coulomb field 
$v_{c}\left[ 
\vec{\mathrm{r}}-\vec{Y}\left( t\right) 
\right]
+ v_{c}\left[ 
\vec{\mathrm{r}}-\vec{X}\left( t\right) 
\right]$, of the atomic cores in the QM, with $\vec{\mathrm{r}}$ its position operator. 
Such a particle is also subject, at large distance form the surface, to the potential of its image charge and of those of the screened nuclei at $\left(\vec{Y},\vec{X}\right)$. 
Denoting with 
$v_{I}\left[
\vec{\mathrm{r}};\vec{Y}\left(t\right) ,
\vec{X}\left( t\right) 
\right]$ 
the sum of these three contributions, we can write the first quantization Hamiltonian of the particle as 
\begin{eqnarray}
h\left( t\right) &=& 
e_{K} + v_{s}\left( \mathrm{z}\right) 
+v_{c}\left[ 
\vec{\mathrm{r}}-\vec{Y}\left( t\right) 
\right] 
+v_{c}\left[ 
\vec{\mathrm{r}}- \vec{X}\left( t\right) 
\right] 
+v_{I}\left[ 
\vec{\mathrm{r}};\vec{Y}\left(t\right) ,\vec{X}\left( t\right) \right] ,               
\nonumber \\
\label{Eq.oeham}
\end{eqnarray}
where atomic units have been used.  
Next, we define the continuous conduction states 
$\left\{ \left| k\right\rangle \right\}$ as eigenstates of the unperturbed surface Hamiltonian 
$h_{s}=e_{K}+v_{s}\left( \mathrm{z}\right)$, 
with spectrum $\left\{ \varepsilon_{k}\right\}$. 
Because of the surface barrier, the corresponding wave functions 
$\left\langle \vec{r}\bigl| k\right\rangle$ drop to zero outside the surface, i.e. at $z>0$, after some atomic distances. 
The adiabatic state 
$\left| a\left( t\right) \right\rangle$ is obtained by translating the stationary state 
$\left| a^{0} \right\rangle$, 
for an electron localized on an atomic nucleus fixed at the surface, to the position of the emitted atom.  
Hence it diagonalizes instantaneously the Hamiltonian 
$h_{a}\left( t\right) =
e_{K}+v_{c}\left[ 
\vec{\mathrm{r}}-\vec{Y} \left( t\right) 
\right]$, 
with a time independent eigenvalue $\varepsilon_{a}^{0}=-I$ (or $\varepsilon_{a}^{0}=A$). 
The atomic wave function depends on $\vec{r}-\vec{Y}$, i.e. 
$\left\langle \vec{r}\bigl|
a\left( t\right) \right\rangle 
=\left\langle \vec{r}-\vec{Y}\left( t\right)\bigl|
a^{0}\right\rangle$. 
As for $\left| m\left( t\right)\right\rangle$, it can be calculated by solving the eigenvalue equation for the QM Hamiltonian
\[ h_{m}\left( t \right)= e_{K}+ 
v_{c}\left[ 
\vec{\mathrm{r}}-\vec{Y}\left( t\right) 
\right]
+v_{c}\left[ 
\vec{\mathrm{r}}-\vec{X}\left( t\right) 
\right] ,
\] viz.
$h_{m}\left( t \right)
\left| m\left( t\right) \right\rangle =
\varepsilon _{m}^{0}\left( t\right) 
\left|m\left( t\right) \right\rangle$, through the LCAO method.

The set of normalized states 
$\bigl\{
\left| k\right\rangle,
\left| a\left(t\right) \right\rangle ,
\left| m\left( t\right) \right\rangle 
\bigr\}$ 
is well defined and can be used as a truncated basis to describe the problem, once the overlap contributions 
$\left\langle a\left( t\right) \bigl|
m\left(t\right) \right\rangle $, 
$\left\langle a\left( t\right) \bigl|k\right\rangle$ 
and 
$\left\langle m\left( t\right) \bigl|k\right\rangle$
have been neglected for all $t$. 
Thus, we can introduce the electron destruction 
$\left( c_{j}\right) _{j\in \{a,m,k\}}$ 
and creation operators 
$\left( c_{i}^{\dagger } \right) _{j\in \{a,m,k\}}$, 
in Schr\"{o}dinger's picture, satisfying algebraic relations 
\begin{equation}
\biggl\{ c_{j} , c_{j^{\prime } } \biggr\} = 0 
\quad ,\quad 
\biggl\{ c_{j} , c_{j^{\prime }}^{\dagger } \biggr\} = \delta _{j,j^{\prime }},
\label{Eq.2.3}
\end{equation}
with $\biggl\{...\biggr\}$ the anticommutation operator and $n_{j}=c_{j}^{\dagger }c_{j}$ electron number operators. 
These may be chosen to be approximately time-independent by an appropriate phase transformation
on $\left| a\left( t\right) \right\rangle$ and 
$\left| m\left(t\right)\right\rangle$ and the many-body system Hamiltonian \cite{Negele} is 

\begin{eqnarray}
&&\mathcal{H}\left( t\right) =
\varepsilon _{a}\left( t\right)n_{a}
+\varepsilon _{m}\left( t\right)n_{m}
+\sum\limits_{k}\varepsilon_{k}n_{k}  
\nonumber \\
&&\qquad \ \ 
+\sum\limits_{k}\biggl\{
V_{ka}\left( t\right) c_{k}^{\dagger}c_{a}
+\mathrm{h.c.}
\biggr\}
+\sum\limits_{k}\biggl\{
V_{km}\left( t\right)c_{k}^{\dagger}c_{m}
+\mathrm{h.c.}
\biggr\}  
\nonumber \\
&&\qquad \ \ 
+\sum\limits_{k,k^{\prime }}
V_{kk^{\prime }}\left( t\right)
c_{k}^{\dagger }c_{k^{\prime }}.  
\label{Eq:2.2}
\end{eqnarray}
Here the atomic and molecular unperturbed energies, $\varepsilon _{0}^{a}$ and $\varepsilon _{0}^{m}$, are shifted in presence of the interactions to 
\begin{eqnarray}  
\varepsilon _{a}\left( t\right) =
\varepsilon _{a}^{0}
+ \left\langle a\left(t\right) \right| 
\biggl\{
v_{c}\left[ 
\mathrm{\vec{r}}-\vec{X}\left( t\right) 
\right] 
+ v_{s}\left( z\right) 
+v_{I}\left[ 
\mathrm{\vec{r}};\vec{Y}\left(t\right) ,
\vec{X}\left( t\right) 
\right] 
\biggr\}
\left| a\left( t\right)\right\rangle  
\nonumber \\
\label{Eq.eat}
\end{eqnarray}
and 
\begin{equation}
\varepsilon _{m}\left( t\right) 
\simeq 
\varepsilon _{m}^{0}\left( t\right)
+ \left\langle m\left( t\right) \right| 
v_{s}\left( \mathrm{z}\right) 
\left| m\left( t\right) \right\rangle ,  
\label{Eq.emt}
\end{equation}
respectively. 
The hopping terms take the form 
\begin{eqnarray}
V_{kj}\left( t\right) & \simeq &
\left\langle k\right| 
\biggl\{v_{c}\left[ 
\vec{\mathrm{r}}-\vec{X}\left( t \right)
\right] 
+v_{c}\left[ 
\vec{\mathrm{r}}-\vec{Y}\left( t \right)
\right] 
\biggr\}
\left| j\left( t\right)\right\rangle ,
\qquad \qquad 
\mathrm{with} \quad 
j\in \left\{ a,m\right\} , 
\nonumber \\
\label{Eq.vkjt}
\end{eqnarray}
and the scattering potential can be written as 
\begin{equation}
V_{kk^{\prime }}\left( t\right) \simeq 
\left\langle k\right| \biggl\{
v_{c} \left[ \vec{\mathrm{r}}-\vec{X}\left( t \right) \right] +v_{c}\left[ \vec{\mathrm{r}}-\vec{Y} \left( t \right) 
\right] 
\biggr\}\left| k^{\prime }\right\rangle .  
\label{Eq.vkk't}
\end{equation}
In Eqs.(\ref{Eq.emt}-\ref{Eq.vkk't}) we have omitted image potential
contributions which act sensibly when the emitted atom is at large distance from the surface where the continuous wave functions vanish and the MO does not contribute. 
Non Anderson-Newns terms enhance the number of electron-hole excitations by processes similar to those reported in Figs.2b and 2c.

Moreover in the reference Hamiltonian we have not allowed direct transitions between the two localized states, regarding terms like 
\[
V_{am}\left( t\right) =
\left\langle a\left( t\right) \right| 
v_{s}\left( \mathrm{z}\right) 
\left| m\left( t\right) \right\rangle =
V_{am}^{\ast}\left( t\right) 
\]
as small perturbations in the energy scale of the hopping interactions. 
We point out that the introduction of the QM contribute in the determination of the quantities $\varepsilon _{a}$ and $V_{ka}$ that, unlike the bare model \cite{Gadzuk,Remy,Nordlander,Kurpick}, depend parametrically on both the outgoing and recoiled atomic trajectories.

Brako and Newns' \textit{slowness approximation} \cite{Brako1,Brako2,Brako3} can be formulated extending its application to the problem of an electron gas probed by \textit{two localized sources}. 
We assume that changes of matrix elements 
$\left( 
\left\langle j\right| h\left| j^{\prime }\right\rangle
\right) _{j,j^{\prime }\in \left\{ k,a,m\right\} }$, 
over the characteristic times of the atomic motion in the QM, are negligible. 
More rigorously we have quasi-stationary dynamics such that 
\begin{equation}
\left\{ 
\begin{array}{c}
\left\langle j\right| h\left| j^{\prime }\right\rangle 
\biggl|_{
\left[ 
\vec{Y }\left( t\right) ,\vec{X}\left( t\right) 
\right] 
}\simeq 
\left\langle j\right| h\left| j^{\prime }\right\rangle 
\biggl|_{
\left[ 
\vec{Y}\left(t+\tau _{Y}\right) ,\vec{X}\left( t\right) 
\right] 
} \\ 
\\ 
\left\langle j\right| h\left| j^{\prime }\right\rangle 
\biggl|_{
\left[ 
\vec{Y }\left( t\right) ,\vec{X}\left( t\right) 
\right]
}\simeq 
\left\langle i\right| h\left| j^{\prime }\right\rangle 
\biggl|_{
\left[ 
\vec{Y}\left(t\right) ,\vec{X}\left( t+\tau _{X}\right) 
\right] 
}
\end{array}.
\right.
\label{Eq.3}
\end{equation}
As we have already observed surface effects rapidly dissipate with respect to Anderson-Newns hopping processes, and a localized molecular electron has a short average life-time in the $\tau _{Y}$-scale. 
Then, for times larger than $\tau _{Y}$, all the terms corresponding to inner excitations may be neglected. 
However while the QM is present many electronic exchanges, see
Figs.2b and 2c, occur that may influence the final charge state of the
sputtered atom.

A very important topic is concerned with the question weather the two discrete states are so much linked that $\left| a\right\rangle$ may transform into $\left| m\right\rangle$, in the first step of the emission process, during the action of the short range atom-atom forces, or if they can be considered distinct and non overlapping. 
In our opinion, the former case can occur when the energies 
$\varepsilon _{a}$ and $\varepsilon _{m}$ are so close, at the time of ejection, viz. $t^{\ast}=0$, that one of the atomic states, localized at $\vec{X}$ and $\vec{Y}$ respectively, in which $\left| m\left(0\right) \right\rangle$ may be decomposed, in the LCAO scheme, is 
$\left|a\left( 0\right) \right\rangle $ itself. 
Anyway, even in this eventuality, we can continue to use the model Hamiltonian of Eq.(\ref{Eq:2.2}), assigning to the state $\left|m\right\rangle$ the role of a discrete atomic state localized on the substrate atom of the QM and transiently forming a MO with the state $\left|a\right\rangle$ in the time interval $0\leq t\leq \tau _{X}$. 
Apart from this conceptual correction no significant qualitative or quantitative changes affect the ionized fraction in the emitted beam, at least within the limits of our approximations. 
Hence, in the following, we shall consider $\left| a\right\rangle $ to be independent from $\left| m\right\rangle$, regarding a more accurate study of correlations between the two discrete
states to future works.

\section{Exact Diagrammatic Solution}

The purpose of our analysis is to find out the fraction of atoms detected at the end of the process with a given charge state. 
In our model this calculation is straightforward once we have determined $c_{a}$ in the final system configuration, when the particle is quite far from the sample and the perturbations are switched off, i.e. $c_{a}\left( \infty \right)$. 
The Hamiltonian (\ref{Eq:2.2}) describes atomic emission only \textit{after} the collision has taken place hence the initial condition should be referred to the time when the atom is at the surface. 
For analytical simplicity we prefer to place the initial time in the remote past and switch on adiabatically the perturbations as in an ion-surface scattering problem. 
It has been argued that this analytical continuation on negative times does not infer the treatment for many systems, such as $Na/W(100)$ \cite{Brako2}. 
We therefore assume that interaction terms obey the following
\[
\left\{ 
\begin{array}{c}
V_{kk^{\prime }}\left( -\infty \right) 
=V_{kk^{\prime }}\left( \infty \right) =0 \\ 
V_{ka}\left( -\infty \right) =
V_{ka}\left( \infty \right) =0 \\ 
V_{km}\left( -\infty \right) 
=V_{km}\left( \infty \right) =0
\end{array}
\right. . 
\]
The neutral fraction of emitted particles at a temperature $T$ is defined as the thermal average of the atomic electron number operator, evolved to the remote future according to Heisenberg's scheme, in the initial equilibrium configuration, or 
\begin{equation}
\mathcal{P}_{0}\left( T\right) \equiv 
\left\langle n_{a}\left( \infty\right) \right\rangle .  
\label{Eq.3.4}
\end{equation}
 Here, spin degrees of freedom have been ignored as Coulomb intra-atomic correlation terms \cite{Brako3,Makoshi,Davidson,Zimny,Shao} do not explicitly appear in the reference Hamiltonian. 
This approximation limits the applicability of the model to very simple ejected beams such as alkalis \cite{Brako1}, and, even in this case, there are non trivial problems at least for the atomic orbital. 
In fact, while the emitted atom moves off surface, the state 
$\left| a \right\rangle$ increases its localization, the hopping mechanism becoming weaker, and approaches the atomic limit, when the exchange of electron with the surface becomes a spin-flip process without charge fluctuation \cite{Brako3}, which cannot be described by a spinless theory. 
On the other hand, the state $\left| m \right\rangle$ acts sensibly on the system only when the emitted atom is in the vicinity of the surface so that its correlation energy may be taken to be infinite, in the scale of hopping interaction strengths, avoiding double occupation of the level. 
Some significative efforts in a better comprehension of spin correlations have been made, in the last decade, on the bare Anderson-Newns model, see for instance Refs.\cite{Davidson,Shao}.
However further studies are needed to discuss the influence of the two-body spin-spin potential in the framework of the present approach. 
Thus, in this analytical derivation, the spin degeneracy can be taken into account only by addiction of a $2$, premultiplying Eq.(\ref{Eq.3.4}), at the end of calculations. 
Complementarily to Eq.(\ref{Eq.3.4}) the ionized fraction can be put in the form 
\begin{equation}
\mathcal{R}_{+}\left( T\right) = 1-\mathcal{P}_{0}\left( T\right)
=\left\langle 
c_{a}\left( \infty \right) c_{a}^{\dagger }\left( \infty\right) \right\rangle .  
\label{Eq.3.4b}
\end{equation}
Our evaluation of (\ref{Eq.3.4}) and (\ref{Eq.3.4b}) are based on the
equations of motion method \cite{Bloss}. 
Electron operators obey the following set of coupled equations 
\begin{equation}
\left\{ 
\begin{array}{c}
i\frac{d}{dt}c_{a}\left( t\right) =
\varepsilon _{a}\left( t\right)c_{a}\left( t\right)
+\sum\limits_{k}V_{ak}\left( t\right) c_{k}\left(t\right) \\ 
\\ 
i\frac{d}{dt}c_{m}\left( t\right) =
\varepsilon _{m}\left( t\right)c_{m}\left( t\right) +\sum\limits_{k}V_{mk}\left( t\right) c_{k}\left(t\right) \\ 
\\ 
i\frac{d}{dt}c_{k}\left( t\right) =
\varepsilon _{k}c_{k}\left( t\right)
+\sum\limits_{k^{\prime }}
V_{kk^{\prime }}\left( t\right) c_{k^{\prime}}\left( t\right) \\ 
\quad \quad \quad \quad \quad \quad 
+V_{ka}\left( t\right) c_{a}\left(t\right) 
+V_{km}\left( t\right) c_{m}\left( t\right)
\end{array}
\right. ,  
\label{Eq.3.7}
\end{equation}
that can be easily written down by using the algebraic relations of Eq.(\ref {Eq.2.3}) extended, in Heisenberg's picture, at all equal times. 
It can be noted that the discrete states $\left| a\right\rangle $, 
$\left|m\right\rangle $ are coupled only through indirect processes, as in Fig.2d.

Eqs.(\ref{Eq.3.7}a-c) can be combined and iterated to give a formal result in terms of time dependent Feynmann graphs: the integral form of (\ref{Eq.3.7}b) is obtained by introducing the unperturbed retarded Green's function, or empty band propagator, of an electron in a state 
$\left| j\right\rangle$, i.e.
\[
G_{j}^{0+}\left( t,t^{\prime }\right) =
-i\theta \left( t-t^{\prime }\right)
e^{-i
\int\limits_{t^{\prime }}^{t}d\tau \varepsilon _{j}\left( \tau \right)
},
\]
as solution of the singular equation 
\[
\left\{ 
i\frac{d}{dt}-\varepsilon _{j}\left( t\right) 
\right\}
G_{j}^{0+}\left( t,t^{\prime }\right) =
\delta \left( t-t^{\prime }\right) .
\]
Then, we have 
\[
c_{m}\left( t\right) =
i G_{m}^{0+}\left( t,-\infty \right)c_{m}
+ \sum\limits_{k}
\int\limits_{-\infty }^{\infty }dt^{\prime}
G_{m}^{0+}\left( t,t^{\prime }\right) 
V_{mk}\left( t^{\prime }\right)
c_{k}\left( t^{\prime }\right) .  
\]
Substituting into Eq.(\ref{Eq.3.7}c), the new equation for $c_{k}\left(t\right)$ reads 
\begin{eqnarray}
&& c_{k}\left( t\right) =
iG_{k}^{0+}(t,-\infty )c_{k}
+\sum\limits_{k^{\prime}}
\int\limits_{-\infty }^{\infty }dt^{\prime }
\int\limits_{-\infty }^{\infty}dt"
G_{k}^{0+}\left( t,t^{\prime }\right) 
W^{m}_{kk^{\prime }}\left( t^{\prime},t"\right) 
c_{k^{\prime }}\left( t"\right)  
\nonumber \\
&&\quad 
+i\int\limits_{-\infty }^{\infty }dt^{\prime }
G_{k}^{0+}\left(t,t^{\prime }\right) 
V_{km}\left( t^{\prime }\right) 
G_{m}^{0+}\left(t^{\prime },-\infty \right) c_{m}
+\int\limits_{-\infty }^{\infty }dt^{\prime}
G_{k}^{0+}\left( t,t^{\prime }\right) 
V_{ka}\left( t^{\prime }\right)
c_{a}\left( t^{\prime }\right) ,  
\nonumber \\
&&  \label{Eq.4.4}
\end{eqnarray}
where surface-induced excitations of Figs.2b and 2c have been enclosed into the \textit{non instantaneous} potential 
\begin{equation}
W^{m}_{kk^{\prime }}\left( t,t^{\prime }\right) \equiv 
V_{kk^{\prime }}\left(t\right) \delta \left( t-t^{\prime }\right)
+V_{km}\left( t\right) G_{m}^{0+}\left( t,t^{\prime }\right)
V_{mk^{\prime }}\left( t^{\prime}\right) .  
\label{Eq.4.5}
\end{equation}
Eq.(\ref{Eq.4.4}) may be substituted into Eq.(\ref{Eq.3.7}a) to obtain 
\begin{eqnarray}
&&\biggl\{
i\frac{d}{dt}-\varepsilon _{a}\left( t\right) 
\biggr\}c_{a}\left(t\right) 
-\int\limits_{-\infty }^{\infty }dt^{\prime }
\Sigma _{a}^{0}\left(t,t^{\prime }\right) c_{a}
\left( t^{\prime }\right)=
i\sum\limits_{k}V_{ak}\left( t\right) 
G_{k}^{0+}\left( t,-\infty \right)c_{k}  
\nonumber \\
&&\quad 
+\sum\limits_{k,k^{\prime }}
\int\limits_{-\infty }^{\infty}dt^{\prime }
\int\limits_{-\infty }^{\infty }dt"
V_{ak}\left( t\right)
G_{k}^{0+}\left( t,t^{\prime }\right) 
W^{m}_{kk^{\prime }}\left( t^{\prime},t"\right) 
c_{k^{\prime }}\left( t"\right)  
\nonumber \\
&&\quad 
+i\sum\limits_{k}
\int\limits_{-\infty }^{\infty }dt^{\prime}
V_{ak}\left( t\right) 
G_{k}^{0+}\left( t,t^{\prime }\right) 
V_{km}\left(t^{\prime }\right) 
G_{m}^{0+}\left( t^{\prime },-\infty \right) c_{m},
\label{Eq.4.6}
\end{eqnarray}
in which $\Sigma _{a}^{0}\left( t,t^{\prime }\right)$ is the retarded
self-energy of the state $\left| a\right\rangle $ in the absence of the QM, 
\begin{equation}
\Sigma _{a}^{0}\left( t,t^{\prime }\right) \equiv
\sum\limits_{k}V_{ak}\left( t\right) 
G_{k}^{0+}\left( t-t^{\prime }\right) 
V_{ka}\left( t^{\prime }\right) .  
\label{Eq.selfaux}
\end{equation}
This term is peculiar to the Anderson-Newns Hamiltonian and responsible for \textit{broadening} and \textit{shift} of the atomic level. 
Surface induced excitations act to renormalize $\Sigma _{a}^{0}$ and $G_{k}^{0+}$; 
after an infinite number of iterations of Eq.(\ref{Eq.4.4}) into Eq.(\ref{Eq.4.6}), we can specify an inner retarded propagator, 
\begin{eqnarray}
&&G_{kk^{\prime }}^{m+}\left( t,t^{\prime }\right) \equiv
G_{k}^{0+}\left( t,t^{\prime }\right) \delta _{k,k^{\prime }}  \nonumber \\
&&\quad 
+\sum\limits_{k"}\int\limits_{-\infty }^{\infty}dt_{1}
\int\limits_{-\infty }^{\infty }dt_{2}
G_{k}^{0+}\left( t,t_{1}\right)
W^{m}_{kk"}\left( t_{1},t_{2}\right) 
G_{k"k^{\prime }}^{m+}\left(t_{2},t^{\prime }\right) ,  
\nonumber
\end{eqnarray}
representing the probability amplitude of a process 
$\left| k^{\prime}\right\rangle \rightarrow \left| k\right\rangle $, 
from time $t^{\prime }$ to time $t$, with multiple empty band scattering from $W^{m}_{kk^{\prime }}$, and we can rewrite Eq.(\ref{Eq.4.6}) as 
\begin{eqnarray}
&&\biggl\{
i\frac{d}{dt}-\varepsilon _{a}\left( t\right) 
\biggr\}c_{a}\left(t\right) 
-\int\limits_{-\infty }^{\infty }dt^{\prime }
\Sigma _{a}\left(t,t^{\prime }\right) 
c_{a}\left( t^{\prime }\right)=
i\sum\limits_{kk^{\prime }}V_{ak}\left( t\right) G_{kk^{\prime}}^{m+}\left( t,-\infty \right) 
c_{k^{\prime }}  
\nonumber \\
&&\quad 
+i\sum\limits_{kk^{\prime }}
\int\limits_{-\infty }^{\infty}dt^{\prime }
V_{ak}\left( t\right) 
G_{kk^{\prime }}^{m+}\left(t,t^{\prime }\right) 
V_{k^{\prime }m}\left( t^{\prime }\right)
G_{m}^{0+}\left( t^{\prime },-\infty \right) c_{m}  
\label{Eq.4.9}
\end{eqnarray}
The \textit{dressed} self-energy graph is then given the form 
\begin{eqnarray} 
\label{Eq.4.10}
\Sigma _{a}\left( t,t^{\prime }\right) \equiv 
\Sigma _{a}^{0}\left(t,t^{\prime }\right) +
\Sigma _{a}^{\mathrm{p}}\left( t,t^{\prime }\right) =
\sum\limits_{kk^{\prime }}V_{ak}\left( t\right) 
G_{kk^{\prime }}^{m+} \left( t,t^{\prime }\right) 
V_{k^{\prime }a}\left( t^{\prime }\right),
\nonumber \\
\end{eqnarray}
where $\Sigma _{a}^{\mathrm{p}}$ is the renormalization term 
\begin{eqnarray}  
\label{Eq.4.12}
\Sigma _{a}^{\mathrm{p}}\left( t,t^{\prime }\right) \equiv
\sum\limits_{kk^{\prime }}
\int\limits_{-\infty }^{\infty}dt_{1}
\int\limits_{-\infty }^{\infty }dt_{2}
V_{ak}\left( t\right)
G_{k}^{0+}\left( t,t_{1}\right) 
T^{m}_{kk^{\prime }}\left( t_{1},t_{2}\right)
G_{k^{\prime }}^{0+}\left( t_{2},t^{\prime }\right) 
V_{k^{\prime }a}\left(t^{\prime }\right) ,  
\nonumber \\
\end{eqnarray}
and 
\begin{eqnarray}
&&T^{m}_{kk^{\prime }}\left( t,t^{\prime }\right) =
W^{m}_{kk^{\prime }}\left(t,t^{\prime }\right)  
\nonumber \\
&\quad &+\sum\limits_{k"}
\int\limits_{-\infty }^{\infty}dt_{1}
\int\limits_{-\infty }^{\infty }dt_{2}
W^{m}_{kk"}\left( t,t_{1}\right)
G_{k"}^{0+}\left( t_{1},t_{2}\right) 
T^{m}_{k"k^{\prime }}\left( t_{2},t^{\prime}\right) ,  
\label{Eq.4.11}
\end{eqnarray}
the retarded transition matrix in the QM potential. 
Integration of (\ref{Eq.4.9}) gives the formal result for the full atomic electron destruction operator 
\[
c_{a}\left( t\right) = 
iG_{a}^{+}\left( t,-\infty \right)c_{a}
+iG_{am}^{+}\left( t,-\infty \right)c_{m}
+i\sum\limits_{k}G_{ak}^{+}\left( t,-\infty \right) c_{k},
\]
in which the full retarded propagators appear 
\begin{equation}
\left\{ 
\begin{array}{c}
\\ 
G_{a}^{+}\left( t,t^{\prime }\right) =
G_{a}^{0+}\left( t,t^{\prime }\right)
+\int\limits_{-\infty }^{\infty }dt_{1}
\int\limits_{-\infty }^{\infty}dt_{2}
G_{a}^{0+}\left( t,t_{1}\right) 
\Sigma _{a}\left( t_{1},t_{2}\right)
G_{a}^{+}\left( t_{2},t^{\prime }\right) \\ 
\\ 
G_{ak}^{+}\left( t,t^{\prime }\right) =
\sum\limits_{k^{\prime}}
\int\limits_{-\infty }^{\infty }dt_{1}
G_{a}^{+}\left( t,t_{1}\right)
V_{ak^{\prime }}\left( t_{1}\right) 
G_{k^{\prime }k}^{m+}\left(t_{1},t^{\prime }\right) \\ 
\\ 
G_{am}^{+}\left( t,t^{\prime }\right) =
\sum\limits_{k}\int\limits_{-\infty}^{\infty }dt_{1}
G_{ak}^{+}\left( t,t_{1}\right) V_{km}\left( t_{1}\right)
G_{m}^{0+}\left( t_{1},t^{\prime }\right)
\end{array}
\right. .  
\label{Eq.4.13}
\end{equation}
Then, the ionization probability is 
\begin{eqnarray}
\mathcal{R}_{+}\left( T\right) &=&\left| 
G_{a}^{+}\left( \infty ,-\infty\right) 
\right| ^{2}\left\langle \bar{n}_{a}\right\rangle 
+\left|
G_{am}^{+}\left( \infty ,-\infty \right) 
\right| ^{2}
\left\langle \bar{n}_{m}\right\rangle 
+\sum\limits_{k}
\left| G_{ak}^{+}\left( \infty ,-\infty\right) \right|^{2}
\left\langle \bar{n}_{k}\right\rangle  
\nonumber \\
&&  \label{Eq.4.17}
\end{eqnarray}
with $\left\langle \bar{n}_{k}\right\rangle $ the unperturbed hole
distribution function 
\[
\left\langle \bar{n}_{k}\right\rangle =
\frac{1}
{e^{
\beta \left( \varepsilon_{f}-\varepsilon _{k}\right) 
}+1},
\]
and 
$\left\langle \bar{n}_{a}\right\rangle$, $\left\langle\bar{n}_{m}\right\rangle$ 
the initial vacancy of the atomic and molecular orbitals. 
Eq.(\ref{Eq.4.17}) generalizes the result of Blandin, Nourtier and
Hone \cite{Blandin} on the basic Anderson-Newns model obtained on the Keldish contour  \cite{Keldish}.

\section{Approximations on the slow dynamics}

A simplified solution for Eq.(\ref{Eq.4.17}) can be obtained by repeatedly performing the following approximations on each term of the perturbation series defining the retarded propagators of Eqs.(\ref{Eq.4.13}):

\begin{enumerate}
\item  
Let us consider the integral 
\[
\mathcal{I}\left( t\right) \equiv 
\sum\limits_{k}
\int\limits_{-\infty}^{t}dt^{\prime }
f_{k}\left( t^{\prime }\right) 
e^{
-i\int\limits_{t^{\prime}}^{t}d\tau 
\left[ 
\varepsilon _{k}-\Omega \left( \tau \right) 
\right] }, 
\]
where $f_{k}$ and is a slowly varying function of time, on the electronic scale, as specified in Sec.3, and 
$\Omega =\varepsilon _{a}$ or $\Omega=\varepsilon _{k}$. 
The oscillating complex exponential in the $t^{\prime}$-integral varies sensibly with $t^{\prime}$ respect to $f_{k}$ so that, in
the summation over the continuous modes ${k}$, it contributes a quasi delta function of $t-t^{\prime }$, at least until the outgoing particle remains in the interaction region with the surface. 
This allows us to hold $f_{k}$ and $\Omega$ constant during integration \cite{Brako1,Brako2} approximating 
\[
\mathcal{I}\left( t\right) \cong 
i\sum\limits_{k} f_{k}\left( t\right) 
G_{k}^{0+}\left[ \Omega \left( t\right) \right] , 
\]
with $\textsc{G}_j^{0+} \left[ \Omega \left( t \right) \right]$ being the Fourier Transform (FT) of the unperturbed retarded Green function 
$G_j^{0+} \left( t \right)$ in the $\varepsilon = \Omega\left( t \right)$-domain, viz.
\[ \textsc{G}_k^{0+} \left[ \Omega \left( t \right) \right] = \frac{1}
{\Omega\left(t\right) - \varepsilon_j \left( t \right) + i \eta} 
\qquad \eta \rightarrow 0^{+}.\]

\item  Each term of the kind 
\[
\mathcal{Y}_{m}\left( t\right) \equiv 
\int\limits_{-\infty }^{t}dt^{\prime}
f_{m}\left( t^{\prime }\right) 
e^{-i\int\limits_{t^{\prime }}^{t} d\tau 
\left[ 
\varepsilon _{m}\left( \tau \right) 
       -\Omega \left( \tau \right) 
\right] }, 
\]
where $f_{m}$ contains the inner hopping matrix elements $V_{km}$, $V_{km}^{\ast }$ simplifies to 
\[
\mathcal{Y}_{m} \left( t\right) \cong 
i\ f_{m}\left( t\right) 
\textsc{G}_{m}^{0+}\left[\Omega \left( t\right) \right] . 
\]
In fact for times $t\leq t^{\prime }\leq \tau_{X}$ the functions $f_{m}$,$\varepsilon _{m}$ and $\Omega $ are slowly varying, i.e. 
$f_{m}\left(t^{\prime }\right) \cong f_{m}\left( t\right)$ 
and 
\[
\int\limits_{t^{\prime }}^{t}d\tau 
\left[ 
\varepsilon _{m}\left( \tau \right) -\Omega \left( \tau \right) 
\right] 
\cong 
\left[ 
\varepsilon_{m}\left( t\right) -\Omega \left( t\right) 
\right] 
\left( t-t^{\prime}\right) , 
\]
while $f_{m}$ switches off to zero for $t^{\prime }\gg \tau_{X}$. 
We point out that for notational convenience we have omitted other quantum indices or time labels on which $f_{k}$ and $f_{m}$ may depend.

\item  The operator 
\[
\mathcal{Q}_{a}\left( t\right) =
\int\limits_{-\infty }^{t}dt^{\prime }
\Sigma_{a}\left( t,t^{\prime }\right) 
c_{a}\left( t^{\prime }\right) , 
\]
may be treated by replacing $c_{a}\left( t^{\prime }\right) $ 
with $iG_{a}^{0-}\left( t^{\prime },t\right) c_{a}\left( t\right) $, with 
$G_{a}^{0-}\left( t^{\prime },t\right)$ the free advanced propagator of the outward atom electron,
i.e. 
\[
\mathcal{Q}_{a}\left( t\right) \cong 
i\int\limits_{-\infty }^{t}dt^{\prime}
\Sigma _{a}\left( t,t^{\prime }\right) 
G_{a}^{0-}\left( t^{\prime},t\right) 
c_{a}\left( t\right) . 
\]
In fact perturbation changes in the full propagator $G_{a}^{-}\left( t^{\prime },t\right)$ are negligible in the neighborhood of $t$ where 
$\Sigma _{a}$, as a function of $t^{\prime }$, reaches its maximum. 
This last is completely equivalent to the semiclassical approximation (SCA) used in Ref.\cite{Shao}.
\end{enumerate}

After some tedious algebraic manipulations, retarded diagrams, introduced in Sec.3, are converted, by means of the approximations introduced above, to instantaneous diagrams. 
In particular the effective QM potential, the $T$-matrix and the renormalized self energy contributions, Eqs.(\ref{Eq.4.5}),(\ref{Eq.4.10}) and (\ref{Eq.4.11}), become

\begin{equation}
\left\{ 
\begin{array}{c}
\\ 
W^{m}_{kk^{\prime }}\left( t,t^{\prime }\right) 
\rightarrow \textsc{W}^{m}_{kk^{\prime}}\left( \varepsilon ,t\right) 
=V_{kk^{\prime }}\left( t\right)
+V_{km}\left( t\right) 
\textsc{G}_{m}^{0+}\left( \varepsilon \right) 
V_{mk}\left(t\right) \\ 
\\ 
T^{m}_{kk^{\prime }}\left( t,t^{\prime }\right) 
\rightarrow \textsc{T}^{m}_{kk^{\prime}}\left( \varepsilon ,t\right) 
=\textsc{W}^{m}_{kk^{\prime }}\left( \varepsilon ,t\right)
+\sum\limits_{k"}\textsc{W}^{m}_{kk"}\left( \varepsilon ,t\right) 
\textsc{G}_{k"}^{0+}\left(\varepsilon \right) 
\textsc{T}^{m}_{k"k^{\prime }}\left( \varepsilon ,t\right) \\ 
\\ 
\Sigma _{a}^{0}\left( t,t^{\prime }\right) 
\rightarrow \Sigma _{a}^{0}\left(\varepsilon ,t\right) =\sum\limits_{k}
\textsc{G}_{k}^{0+}\left( \varepsilon \right)
\left| V_{ak}\left( t\right) \right| ^{2} \\ 
\\ 
\Sigma _{a}^{\mathrm{p}}\left( t,t^{\prime }\right) 
\rightarrow \Sigma _{a}^{\mathrm{p}}\left( \varepsilon ,t\right) =\sum\limits_{k,k^{\prime}}V_{ak}\left( t\right) 
\textsc{G}_{k}^{0+}\left( \varepsilon \right) 
\textsc{T}^{m}_{kk^{\prime}}\left( \varepsilon ,t\right) 
\textsc{G}_{k^{\prime }}^{0+}\left( \varepsilon\right) 
V_{k^{\prime }a}\left( t\right)
\end{array}
\right. ,  
\label{Eq.instantaneous}
\end{equation}
where $\textsc{T}^{m}_{kk^{\prime }}\left( \varepsilon ,t\right)$ is a Brako and Newns instantaneous $T$-matrix \cite{Brako1} and the real and imaginary parts of $\Sigma _{a}\left( \varepsilon ,t\right)$ 
give respectively the instantaneous \textit{shift} and \textit{broadening} (see Eq.(\ref{Eq:2.1})) of the atomic level in interaction with the QM. 
Eq.(\ref{Eq.4.9}) is transformed into the integrable form 
\begin{eqnarray}
&&\biggl\{
i\frac{d}{dt}-\varepsilon _{a}\left( t\right) 
-\Sigma _{a}\left(t\right) 
\biggr\}c_{a}\left( t\right) =  
\nonumber \\
&&\qquad \qquad 
\lim_{t_{0}\rightarrow -\infty }
\biggl\{
\sum\limits_{k,k^{\prime }}
V_{ak}\left( t\right) e^{-i\varepsilon _{k}t}
\left[ 
\delta_{k,k^{\prime }}
-i\int\limits_{-\infty }^{t}dt^{\prime }
e^{i\left(
\varepsilon _{k}-\varepsilon _{k^{\prime }}
\right) t^{\prime }}
\textsc{T}^{m}_{kk^{\prime}}\left( \varepsilon _{k^{\prime }},t^{\prime }\right)
\right]
e^{i\varepsilon_{k^{\prime }}t_{0}}c_{k^{\prime }}  
\nonumber \\
&&\qquad \qquad 
-i\sum\limits_{k}V_{ak}\left( t\right) 
e^{-i\varepsilon_{k}t}
\int\limits_{-\infty }^{t}dt^{\prime }
e^{i\left[ 
\varepsilon_{k}-\varepsilon _{m}\left( t\right) 
\right] t^{\prime }}
\textsc{T}_{km}\left[
\varepsilon _{m}\left( t^{\prime }\right) ,t^{\prime }
\right]
e^{i \varepsilon _{m}\left( t\right) t_{0}}c_{m}
\biggr\},  
\label{Eq.5.16}
\end{eqnarray}
with $\Sigma _{a}\left( t\right)$ standing for 
$\Sigma _{a}\left[\varepsilon _{a}\left( t\right) ,t\right]$ 
and 
\begin{equation}
\textsc{T}_{km}\left[ 
\varepsilon _{m}\left( t\right) ,t
\right] =
\sum\limits_{k^{\prime }}
\textsc{T}^{m}_{kk^{\prime }}\left[ \varepsilon _{m}\left( t\right) ,t\right]
\textsc{G}_{k^{\prime }}^{0+}\left[ \varepsilon _{m}\left( t\right) \right]
V_{k^{\prime }m}\left( t\right) .  
\label{Eq.TTM}
\end{equation}
Eq.(\ref{Eq.5.16}) can be solved by introducing the 
\textit{semiclassical propagator} of an electron in the state 
$\left| a\right\rangle $,i.e. 
\[
G_{a}^{\mathrm{sc}+}\left( t,t^{\prime }\right) =
-i\Theta \left( t-t^{\prime}\right) 
e^{-i\int\limits_{t^{\prime }}^{t}d\tau 
\left[ 
\varepsilon_{a}\left( \tau \right) +\Sigma _{a}\left( \tau \right) 
\right] 
}.
\]
Hence the \textit{slow solution} for the Heisenberg atomic electron operator is 

\begin{eqnarray}
c_{a}\left( t\right) &=&
\lim_{t_{0}\rightarrow -\infty}
\sum\limits_{kk^{\prime }}
\int\limits_{-\infty }^{t}dt^{\prime}
e^{-i\int\limits_{t^{\prime }}^{t}d\tau 
\left[ 
\varepsilon _{a}\left( \tau\right) 
+\Sigma _{a}\left( \tau \right) 
\right] }
V_{ak}\left( t^{\prime}\right) 
e^{-i\varepsilon _{k}t^{\prime }}  
\nonumber \\
&&\times 
\biggr[ \delta _{k,k^{\prime }}-
i\int\limits_{-\infty }^{t^{\prime}}dt"
e^{i \left( 
\varepsilon _{k}-\varepsilon _{k^{\prime }}
\right)t"}
\textsc{T}^{m}_{kk^{\prime }} \left( \varepsilon _{k^{\prime }},t"\right) 
\biggr] 
e^{i\varepsilon _{k^{\prime }}t_{0}}c_{k^{\prime }},  
\nonumber
\end{eqnarray}

where we have ignored the terms containing information on the initial
vacancy, referred to the remote past, of the atomic and molecular
orbitals.  
These assumptions fit the case understudy well. 
In fact while atoms move away from the surface so many exchanges occur with the conduction band that the single transition processes 
$\left| a\right\rangle \rightarrow \left| a\right\rangle $ and 
$\left| m\right\rangle \rightarrow \left| a\right\rangle $ 
may be ignored. 
From Eq.(\ref{Eq.3.4b}) we have the ionized flux as function of instantaneous terms that can be appropriately modeled according to the outgoing ion and surface motion 
\begin{eqnarray}
\mathcal{R}_{+}\left( T\right) &=&
\sum\limits_{k}\left| 
\int\limits_{-\infty}^{\infty }dt
\sum\limits_{k^{\prime }}
e^{i\int\limits_{t}^{\infty }
d\tau \left[ 
\varepsilon _{k^{\prime }}-\varepsilon _{a}\left( \tau \right)
-\Sigma _{a}(\tau )
\right] }
V_{ak^{\prime }}\left( t\right) \right. 
\nonumber \\
&&\qquad \quad \times \left. \biggl[
\delta _{k,k^{\prime}}
-i\int\limits_{-\infty }^{t}dt^{\prime }
e^{i\left( 
\varepsilon _{k^{\prime}}-\varepsilon _{k}
\right) t^{\prime }}
\textsc{T}^{m}_{k^{\prime }k}\left( 
\varepsilon_{k},t^{\prime }
\right) 
\biggr]
\right| ^{2}
\left\langle \bar{n}_{k}\right\rangle  
\label{Eq.6.1}
\end{eqnarray}
Surface effects are included within the instantaneous multiple transition term $\textsc{T}^{m}_{k^{\prime }k}$ acting to renormalize the instantaneous self energy and producing an additional term within the square modulus in the r.h.s. of the previous equation. 
If we switch off the QM potential, letting 
$\textsc{W}^{m}_{kk^{\prime }}\rightarrow 0$ and hence 
$\textsc{T}^{m}_{kk^{\prime }}\rightarrow 0$, we reattain the well known result \cite{Blandin,Brako1,Lang1} 
\begin{equation}
\mathcal{R}_{+}\left( T\right) =
\sum\limits_{k}\left| 
\int\limits_{-\infty}^{\infty }dt 
e^{
i\int\limits_{t}^{\infty }d\tau 
\left[ 
\varepsilon_{k}-\varepsilon _{a}\left( \tau \right) 
-\Sigma _{a}^{0}\left( \tau \right) 
\right] }
V_{ak}\left( t\right) 
\right| ^{2}
\left\langle \bar{n}_{k}\right\rangle .  
\label{Eq.7.2}
\end{equation}
Our analytical investigations focus on the \textit{non adiabatic} ionized flux at very low emission energies, i.e. 
$1\preceq E\preceq 20$ \textrm{eV}. 
Non adiabaticity is achieved by constraing the imaginary part of the
instantaneous self energy to vanish asymptotically more rapidly than $1/t$ \cite{Blandin}. 
In addition we believe that, within the energy range considered, only the electron-hole excitation matrix element in the r.h.s of Eq.(\ref{Eq.6.1}) contributes, in other words continuous electrons are strongly perturbed by $\textsc{W}^{m}_{kk^{\prime }}$ undergoing multiple scattering processes before hopping into the atomic level. 
For mathematical, simplicity we observe that both $V_{km}$ and $V_{kk^{\prime }}$ are much more sensitive on $\vec{X}$, so that we may neglect the $\vec{Y}$-dependence, or assume that QM potentials are effectively localized on the substrate atom.
The unperturbed substrate may be approximated by a free electron model with $k$ representing the eigenvalues of the one-electron wave vector operator, i.e. $k=\vec{k}$. 
Indeed $\vec{k}$ should be a complex vector, taking into account
the decaying part of continuous wave functions outside the surface, yet we neglect its imaginary part since the QM interactions are peaked in the space region where a metal electron can be considered to be in a plane-wave state, therefore this change does not modify substantially our calculations.
We can now perform a phase transformation on the instantaneous $T$-matrix of Eq.(\ref{Eq.instantaneous}b) expressing the ionization probability in terms of a reduced $\mathcal{T}$-matrix localized at the surface 
\begin{equation}
\mathcal{T}^{m}_{\vec{k}\vec{k^{\prime }}}
\left( {\varepsilon }_{k^{\prime}},t\right) =
e^{i\left( \vec{k}-\vec{k^{\prime }}\right) 
\cdot \vec{X}\left(t\right) }
\textsc{T}^{m}_{\vec{k}\vec{k^{\prime }}}
\left( \varepsilon _{k^{\prime}},t\right) .  
\nonumber
\end{equation}
Next we assume separability of $V_{ka}$ between $k$ and t 
dependence {\cite{Anderson}-\cite{Falcone}}, 
$V_{ka}\left( t\right) =
 V_{\left| \vec{k}\right|} \mathrm{u}_{a}\left( t\right)$, 
where $\mathrm{u}_{a}$ is entirely determined by the atomic emission trajectory, ignoring the correction due to the substrate atom, and, referring to the simplest case, we adopt a rigid trajectory parameterization. 
We replace $\vec{Y}$ by its average 
$\left\langle \vec{Y}\right\rangle
=\left\langle \mathrm{\vec{v}}_{\mathrm{T}}\right\rangle t$ 
so that 
$\mathrm{u}_{a}\equiv 
\exp \left( -\frac{\gamma}{2}\mathrm{v}t\right)$, 
in which $\mathrm{v}$ is the component of the
average outgoing velocity $\mathrm{\vec{v}}_{T}$ in the orthogonal direction to the surface plane \cite{Lang3,Wunnik} and 
$\left( \gamma \mathrm{v}\right) ^{-1}$ a characteristic time of the order of $\tau _{Y}$. 
Moreover, working under resonance conditions of $\varepsilon _{a}$ and $\varepsilon_{m}$ with $\varepsilon _{f}$, the single particle density of states $\rho $ as well as the stationary part of the Anderson-Newns hopping potential can be fixed at the Fermi surface {\cite{Blandin,Brako1}} and $\Sigma _{a}^{0}$ becomes a purely imaginary quantity, i.e. 
$\Sigma _{a}^{0}\left( t\right)=
-i\Delta _{a}\exp \left( -\gamma \mathrm{v}t\right) $.

Furthermore we consider scattering transitions induced by the reduced inner potential of Eq.(\ref{Eq.instantaneous}a) to be isotropic, see 
Eqs.(\ref{Eq.vkjt}), (\ref{Eq.vkk't}), and we use the approximated relation 
\[
\frac{1}{2\pi \rho }
\sum\limits_{\vec{k}^{\prime }}
e^{-i
\varepsilon_{k^{\prime }}
\left( t-t^{\prime }\right) 
-i \vec{k^{\prime }}\cdot \vec{X}\left( t\right) 
} 
\mathcal{T}^{m}_{ \left| \vec{k^{\prime }}\right| 
\left| \vec{k}\right| }
\left( \varepsilon _{k},t^{\prime }\right) \cong 
\frac{
\sin \left[k_{f} \left| \vec{X}\left( t\right) \right| \right] }{
k_{f}\left| \vec{X}\left( t\right) \right| 
}
\mathcal{T}^{m}_{k_{f}\left| \vec{k}\right| }
\left(\varepsilon _{k},t\right) 
\delta \left( t-t^{\prime }\right) , 
\]
with $k_{f}$ the Fermi wave vector, obtaining
\begin{eqnarray}  
\label{Eq.9.1}
\mathcal{R}_{+}\left( T\right) &\cong &
4\pi \rho ^{2}\Delta _{a}
\int \frac{d\varepsilon }
{e^{\beta \left( \varepsilon _{f}-\varepsilon \right) }+1}
\left| \int\limits_{-\infty }^{\infty }dt
\frac{\sin \left[ 
k_{f}\left| \vec{X }\left( t\right) \right| 
\right] }{k_{f}\left| \vec{X}\left( t\right)\right| }
e^{-\frac{\gamma }{2}\mathrm{v}t}
\right.  
\nonumber \\
&&\qquad \qquad \qquad \qquad \qquad \qquad 
\times \left.
e^{i\int_{t}^{\infty }d\tau 
\left[ \varepsilon -\varepsilon _{a}\left( \tau\right) 
-\Sigma _{a}^{0+}\left( \tau \right) 
\right] }
\mathcal{T}^{m}_{k_{f}\left| \vec{k}_{\varepsilon }\right| }
\left( \varepsilon ,t\right)
\right| ^{2},  
\nonumber \\
\end{eqnarray}
in which the correlation term 
\begin{equation}
\Sigma _{a}^{\mathrm{p}}\left( t\right) 
\cong -\pi \rho \Delta _{a}
\left\{ 
\frac{\sin \left[ 
k_{f}\left| \vec{X}\left( t\right) \right| \right] 
}{k_{f}\left| \vec{X}\left( t\right) \right| }
\right\} ^{2}
e^{-\gamma \mathrm{v}t}
\mathcal{T}^{m}_{k_{f}k_{f}}
\left[ \varepsilon _{a}\left( t\right) ,t\right] ,
\label{Eq.pse}
\end{equation}
has been omitted. 
We have thus selected charge exchanges, mediated by the
Fermi level, of the type reported in Fig.2d. 
The main part of the process takes place over the $\tau _{X}$-scale. 
In the hypothesis 
$\frac{1}{\gamma \mathrm{v}}\gg \tau _{X}$ 
we can replace the factor 
\[
F\left( t\right) =
e^{-\frac{\gamma }{2}\mathrm{v}t}
\exp \left( 
-\frac{\Delta_{a}}{\gamma \mathrm{v}}
e^{-\gamma \mathrm{v}t}
\right) ,
\]
within the t-integral in Eq.(\ref{Eq.9.1}), with its maximum value $F_{\mathrm{max}}\sim 
\sqrt{\frac{\gamma \mathrm{v}}{5\Delta _{a}}}$. 
In addition $\varepsilon _{a}\left( t\right) $ is evaluated at the time $t_{0}=
\ln \left( 2\Delta _{a}/\gamma \mathrm{v}\right) 
/\gamma \mathrm{v}$,
when $F\left( t_{0}\right) =F_{\mathrm{max}}$. 
Consequently we have the finite temperature generalization of Eq.(16) of Ref.\cite{Sroubek3}(apart from a factor 4 due to a typographical error) 
\begin{equation}
\mathcal{R}_{+}\left( T\right) =
\frac{4\pi }{5}\gamma \mathrm{v}
\bar{N}_{T}\left( \varepsilon _{a},\varepsilon _{f}\right) ,  \label{Eq.ionprobNa}
\end{equation}
where 
\begin{eqnarray}
\bar{N}_{T}\left( \varepsilon _{a},\varepsilon _{f}\right) &=&
\int \frac{d\varepsilon}
{e^{\beta \left( \varepsilon _{f}-\varepsilon \right) }+1}
\left| 
\rho \int\limits_{-\infty }^{\infty }dt
\frac{
\sin \left[ k_{f}\left| \vec{X}\left( t\right) \right| \right] 
}{k_{f} \left| 
\vec{X}\left( t\right)
\right| }
\mathcal{T}^{m}_{k_{f} \left| \vec{k}_{\varepsilon }\right| }
\left( \varepsilon ,t \right) 
e^{-i\left( \varepsilon -\varepsilon _{a}\right)t}
\right| ^{2},  
\nonumber \\
\label{Eq.ionprobaux}
\end{eqnarray}
is the number of holes in thermal equilibrium excited to the Fermi level by the effective potential $\textsc{W}^{m}_{kk^{\prime }}$. 
Our problem has been reduced to a time parameterization of the instantaneous $\mathcal{T}$-matrix. 
We note that the ionized fraction is directly proportional to the atom velocity in contrast to the exponential dependence on 
$\mathrm{v}^{-1}$ obtained at higher kinetic energies and low temperatures \cite{Blandin}. 
This may be taken as an a posteriori justification of the approximation that let us neglect the first term in the r.h.s. of Eq.(\ref{Eq.6.1}).

A simple analytical formula can be obtained if in $\textsc{W}^{m}_{kk^{\prime }}$ we retain only the hopping term $V_{km}$ due to the promoted MO. 
Hence the series (\ref{Eq.instantaneous}b) can be summed exactly to yield
\begin{equation}
\mathcal{T}^{m}_{k_{f}\left| \vec{k}_{\varepsilon }\right| }
\left( \varepsilon,t\right) =
\frac{
\mathcal{V}_{k_{f}m}\left( t\right) 
\textsc{G}_{m}^{0+}\left(\varepsilon \right) 
\mathcal{V}_{m\left| \vec{k}_{\varepsilon }\right|}\left( t\right) }{
1-\textsc{G}_{m}^{0+}\left( \varepsilon \right) 
\Sigma _{m}\left(\varepsilon ,t\right) 
},  
\label{Eq.Birillo}
\end{equation}
where $\Sigma _{m}$ is the instantaneous retarded self-energy of an electron in $\left| m\right\rangle $. 
We observe that $\vec{X}\left( t\right) $ may be replaced with its average value, for $0\leq t\leq \tau _{X}$, when the
t-integrand in Eq.(\ref{Eq.ionprobaux}) is sharply peaked. 
In the same interval {$\varepsilon $}$_{m}$ depends linearly on time, with a slope $b$ depending on its position relative to $\varepsilon _{f}$ (see Fig.1).
Indicating with $t_{1}$ the crossing time, we set 
\begin{equation}
\varepsilon _{m}\left( t\right) =
\varepsilon _{f}+
\mathrm{sgn}\left(
\varepsilon _{af}
\right) b\left( t-t_{1}\right) ,  
\label{Eq.em}
\end{equation}
with $\varepsilon _{af}=\varepsilon _{a}-\varepsilon _{f}$. 
Moreover $\mathcal{V}_{\left| \vec{k}\right| m}$ can be evaluated at the average 
$X = \left\langle 
\left| \vec{X}\left( t\right) \right| 
\right\rangle _{t\in\lbrack 0,\tau _{X}]}$.
Thus we can use the expression 
\begin{equation}
\mathcal{T}^{m}_{k_{f}\left| \vec{k}_{\varepsilon }\right| }
\left( \varepsilon,t\right) =
\frac{1}{\pi \rho }
\frac{\Delta _{m}}
{\varepsilon -\varepsilon_{f}-
\mathrm{sgn}\left( \varepsilon _{af}\right) 
b\left( t-t_{1}\right)
+i\Delta _{m}
},  
\label{Eq.Bbirillo}
\end{equation}
with $\Delta _{m}=
\pi \rho \left| \mathcal{V}_{k_{f}m}\right| ^{2}$ 
being the maximum broadening of the promoted MO, into Eq.(\ref{Eq.ionprobaux}) and continue to integrate from $-\infty $ to $\infty $ for analytical simplicity. 
Employing the FT of (\ref{Eq.Bbirillo}) in the 
$\left(
\varepsilon -\varepsilon _{a}\right) 
$-domain we find 
\begin{eqnarray}
\bar{N}_{T}\left( \varepsilon _{af}\right) &=&
\frac{4\Delta _{m}^{2}}{b^{2}}
\left( 
\frac{\sin k_{f}X}{k_{f}X}
\right) ^{2}
\biggl\{
\Theta \left(\varepsilon _{af}\right) 
\int_{-\infty }^{\varepsilon _{af}}\frac{d\xi 
e^{
- \frac{2\Delta _{m}}{b}\left( \varepsilon _{af}-\xi \right) 
}}{e^{-\beta \xi}+1}  \nonumber \\
&&\qquad \qquad \qquad \qquad \quad 
+\Theta \left( -\varepsilon _{af}\right)
\int_{\varepsilon _{af}}^{\infty }
\frac{d\xi 
e^{-\frac{2\Delta _{m}}{b}
\left( \xi -\varepsilon _{af}\right) }}
{e^{-\beta \xi }+1}
\biggr\},
\label{Important}
\end{eqnarray}
having performed the change of the variable 
$\xi =\varepsilon -\varepsilon_{f}$. 
Eq.(\ref{Important}) can be integrated analytically, neglecting
contributions of the order of 
$e^{-\beta \left| \varepsilon _{af}\right| }$,
to obtain 
\begin{eqnarray}
\mathcal{R}_{+}\left( T\right) &=&
\Theta \left( \varepsilon _{af}\right) - 
\mathrm{sgn}\left( \varepsilon _{af}\right) 
\left( \frac{\sin k_{f}X}{k_{f}X}\right) ^{2}
\frac{16\pi ^{2} \Delta _{m}^{2} \gamma \mathrm{v}}
{5b^{2}\beta \sin \left( 2\pi \Delta _{m}/b\beta \right) }
e^{-2\frac{\Delta _{m}}{b} \left| \varepsilon _{af}\right| }.  \nonumber \\
\label{Eq.10.4}
\end{eqnarray}
Such an approximation is referred to the case of a soft promotion, with $b/2\Delta _{m}\sim 0.1$ \textrm{eV} for typical values of 
$\left| \varepsilon_{af}\right|$ of the order of $1$ eV 
and temperatures up to $1000$ K. 
Moreover when thermal effects become negligible, i.e. 
$\Delta _{m}/b\beta \ll 1$, we obtain \cite{Anto} the generalization of Eq.(47) of Ref.\cite{Sroubek3} 
\begin{equation}
\mathcal{R}_{+}\cong \Theta \left( \varepsilon _{af}\right) -\mathrm{sgn} \left( \varepsilon _{af}\right) 
\frac{8\pi \gamma \mathrm{v}\Delta _{m}}{5b}
\left( 
\frac{\sin k_{f}X}{k_{f}X}
\right) ^{2}
e^{-2 \frac{\Delta _{m}}{b}
\left| \varepsilon _{af}\right| 
},  
\label{Eq.10.5}
\end{equation}
including the case $\varepsilon _{af}>0$. Eqs.(\ref{Eq.ionprobNa}),(\ref{Eq.10.5}) 
are to be compared with the results of Brako and Newns 
(Eq.(34)of Ref.\cite{Brako1}) 
\begin{equation}
\mathcal{R}_{+}\left( T\right) =
\frac{1}{\gamma \mathrm{v}}
\int\limits_{-\infty }^{\infty }d\xi 
\frac{
\mathrm{sech} \left\{ 
\pi \left( \xi -\varepsilon _{af} \right) 
/\gamma \mathrm{v}
\right\} }{e^{-\beta \xi }+1},
\label{Eq.34BN}
\end{equation}
holding in the $1/\beta \gg \gamma v/\pi$-limit, and Blandin, Nourtier and Hone (Eq.(61) of Ref.\cite{Blandin}) 
\begin{equation}
\mathcal{R}_{+}\cong \Theta \left( \varepsilon _{af}\right) -\frac{2}{\pi } \mathrm{sgn}\left( \varepsilon _{af}\right) 
e^{-\pi \left| \varepsilon_{af}\right| /\gamma \mathrm{v}}.  \label{Eq.61BNH}
\end{equation}
These can be obtained from Eq.(\ref{Eq.7.2}) under the same assumptions that lead to Eq(\ref{Eq.ionprobNa}).

\section{Comparison of Ionization Probabilities and Conclusions}

In deriving a rough estimation of the fundamental parameters of the problem we note that the characteristic energies 
$\varepsilon _{0}=\gamma \mathrm{v}/\pi $ and 
$\varepsilon _{1}=b/2\Delta _{m}$ are of the order of 
$1/\tau_{Y} $ and $1/\tau _{X}$ respectively. 
Thus, under the working hypotheses of this paper 
$\left| \varepsilon _{af}\right| $ must be larger than both $\varepsilon _{1}$ and $\varepsilon _{0}$ while we should also have $\varepsilon _{1}\gg \varepsilon _{0}$. 
The evaluation of the average outward velocity $\mathrm{v}$ requires further consideration: 
it has been well established that when the sputtered ion is in the interaction region of the surface its actual velocity is higher than its final measured value $\mathrm{v}_{f}$ owing to the attractive potential. 
In particular this should be true in the low emission energy range under study. 
According to Lang's model \cite{Lang2,Lang3}, the binding potential of the outgoing atom, of mass $m$, with respect to a substrate atom of mass $M$,see Fig.3, can be written in a Morse form 
\[
\mathcal{U}\left( s\right) = \xi _{b}\left[ 
1 - e^{\alpha \left( s-s_{b}\right)}
\right] -\xi _{b}, 
\]
with $s$ the internuclear separation, viz. 
$s=\left|\vec{Y}-\vec{X}\right|$, $s_{b}$ the equilibrium distance, $1/\alpha$ a characteristic interaction length and $\xi _{b}$ the binding energy. 
At $t=0$, when the emission takes place, the substrate atom receives a sudden impulse from other substrate atoms in the collision cascade and the evolution of the emitted atom along the normal surface direction may be calculated by Energy and Momentum conservation. 
We have calculated the average velocity of the outward motion
as function of $\mathrm{v}_{f}$ 
\begin{equation}
\mathrm{v}=\frac{1}{2}\left( \mathrm{v}_{f}+
\frac{\mathrm{v}_{b}^{2}}{\mathrm{v}_{f}}
\right),  
\label{Eq.avvel}
\end{equation}
in which $\mathrm{v}_{b}^{2}=2\frac{M+m}{Mm}\xi _{b}$. 
Note that in this very simple model other substrate atoms do not interact with the QM. 
For typical binding energies of Alkalis and Copper on clean surfaces between $5$ and $20$ eV we have $10^{5}\preceq \mathrm{v}_{b}\preceq 10^{6}$ cm/s. 
Since $\gamma ^{-1}$ is a characteristic length of the
Anderson-Newns interaction, of the order of $1$ \AA, 
$\varepsilon_{0}$ can take values from $10^{-3}$ to $10^{-2}$ eV at emission energies from $1$ to $10$ eV. 
Consequently 
$0.01\preceq \varepsilon_{1}\preceq 0.1$ eV. 
The majority of the metals of interest have Fermi wave vectors in the range $0.5\preceq k_{f}\preceq 2$ \AA $^{-1}$and the average position of the substrate atom $X$, in the interaction region, must be of the order of a few surface layers, i.e. $1\preceq X\preceq 5$ \AA. 
More quantitatively, in Fig.5 we have reported a graphic comparison of
Eqs.(\ref{Eq.10.5}),(\ref{Eq.61BNH}) as functions of 
$\varepsilon _{af}$ with the parameters of an alkali sputtering experiment at emission energies of $5$ eV. 
Despite the two results exhibiting the same dependence on 
$\varepsilon _{af}$, Eq.(\ref{Eq.61BNH}) seems to excessively
underestimate the ionized flux for $\varepsilon _{af}\leq -0.2$ eV,
quickly approaching the adiabatic regime. 
The experimentally determined ionization coefficient  of Cu$^{+}$ sputtering from clean Cu\cite{Sroubek2} is plotted in Fig.6 as function of the emission velocity, $\mathrm{v}_{f}$. 
At energies below $10$ eV the velocity behavior of Eq.(\ref{Eq.10.5}) is in good agreement with data while at higher energies, above $30$ eV, the basic result of Eq.( \ref{Eq.61BNH}) is still valid.

Inclusion of temperatures, by Eqs.(\ref{Important}) and (\ref{Eq.10.4}),allows us to conclude that, at least in the limit of our model, surface effects are somewhat less sensitive to thermal interactions than Anderson-Newns' hopping mechanism. 
In Fig.7 we have reported the functional dependence of $\mathcal{R}_{+}$ on $T$, for an emission energy of the order of $1$ eV, a binding energy of $1.2$ eV, i.e. $\varepsilon_{0}=2.7 \times 10^{-3}$ eV,
 by numerically integrating Eq.(\ref{Important}), the gray line corresponding to the analytical solution of Eq.(\ref{Eq.10.4}). 
The energy separation $\varepsilon _{af}$ has been set to $-0.4$ eV, typical of Cu$^{+}$/Cu sputtering, and the range of validity of the analytical result (\ref{Eq.10.4}) increases, from $T \preceq 500$ K to $T \preceq 850$ K, with increasing $\varepsilon_{1}$ (it must be noted anyway that when $\varepsilon_{1}=0.4$ eV or 
$\varepsilon_{1} = 0.8$ eV the slowness approximation can be hardly applied, $\tau_{Y}$ being of the order of 
$\left|\varepsilon_{af}\right|^{-1}$). 
As shown in Fig.8, Eq.(\ref{Eq.61BNH}) varies over several orders of magnitude with increasing temperature while Eq.(\ref{Eq.10.4}) undergoes small variations, within one order of magnitude (see also Fig.7). 
On the other hand the latter is strongly influenced by changes in the characteristic energy $\varepsilon _{1}$.
We note that, at high temperatures, above $600$ K, $\mathcal{R}_{+}$ is mainly determined by the basic Anderson-Newns model, even in the low $\mathrm{v}_f$-region, while in the interval $200\leq T\leq 500$ K, for $0.08\leq \varepsilon _{1}\leq 0.8$ eV, QM interactions are competitive with the former. 
It should also be observed that the limit of validity of Eq.(\ref{Eq.34BN}) is not strictly respected below $300$ K, which would mean that the result form the basic mechanism, reported in the above mentioned figure, is over-estimated. 
On the other hand our result Eq.(\ref{Important}) can be applied at all temperatures, provided that resonant charge-exchange through the MO is the leading process. 
Inclusion of thermal interactions attests that the two results here discussed are not so dramatically different as the zero temperature theory would suggest. 
In this prospective, it seems an important task for future
studies to analyze more quantitatively correlations between these behaviors, devoting much attention to all the contributions appearing in 
Eq.(\ref{Eq.6.1}) and, particularly, to the perturbed self-energy of Eq.(\ref{Eq.pse}).

To conclude we observe that finding a power law dependence of $\mathcal{R}_{+}$ on $\mathrm{v}_{f}$, observed in much experimental work \cite {Hart,Brizzolara}, is still an open theoretical problem. 
From the scarce contribution found in literature it looks evident that the sole resonant interaction of the atomic state with metal electrons is not sufficient to obtain such a behavior. 
In Refs.\cite{Sroubek1,Sroubek2} a solution of the
form $\mathcal{R}_{+}\propto \mathrm{v}_{f}^{\delta }$ has been obtained through the introduction of an effective substrate temperature into a semiclassical master equation for $\left\langle n_{a}\right\rangle $, while in Refs.\cite{Sroubek3,Anto} the result 
$\mathcal{R}_{+}\propto \mathrm{v}_{f}$ 
has been proposed as solution of an Anderson-Newns Hamiltonian in the screened scattering potential of a moving source localized at atomic distances from the surface. 
These papers have provided the grounds from the
present analysis, in which non Anderson interaction have been more accurately modeled by the transient QM. 
A different approach  \cite{Veksler}, taking into consideration the screening of the nucleus field of a sputtered atom by electrons simultaneously tunneling in its vicinity through the
potential barrier (see Figs.4) , has yield a velocity dependence of the form  
$R_{+}=H\mathrm{v}_{f}\left( 1+B\mathrm{v}_{f}\right) $. 
In the same paper, the author has pointed out that the formula does not hold at high emission energies. 
In the present work, we have explained the different behavior of
Eqs.(\ref{Important}), (\ref{Eq.61BNH}) and Eqs.(\ref{Eq.10.5}),(\ref{Eq.34BN}), in the framework of the same model, as depending on the energy received by the sputtered particle in the collision cascade. 
On the contrary, in Ref.\cite{Veksler}, the two results are presented in relation to the use of two different approaches, the simultaneous resonant tunneling and the
Anderson-Newns, the second of which is put in doubt because of the low $\mathrm{v}_{f}$ features of $\mathcal{R}_{+}$. 
Our study attests the correctness of this latter, in its bare form, providing an implementation that may explain some incongruencies with experiments in the critical low $\mathrm{v}_{f}$ region at low-medium temperatures.

Finally, it is worth to mention that a Hamiltonian resembling Eq.(\ref{Eq:2.2}), with two nearly stationary and directly interacting atomic states, has been used in a few different contexts. 
In Ref.\cite{Kato} a similar model has been applied to the study of resonant charge transfer of scattering
proton from an adsorbate-covered metal surface. 
There the authors have modeled different atomic dynamics, regarding the substrate atom as fixed, thus allowing sensible overlapping between the atomic states, and have
calculated the ionization coefficient as function of the characteristic time of emission, here denoted by $\tau _{Y}$. 
The analytical derivation has been made on an effective Anderson-Newns Hamiltonian obtained by a canonical
transformation, thus formally eliminating the second discrete state, and compared to simulations. 
It can be easily shown that the ionization probability of Ref.\cite{Kato}, once coupling terms between the discrete
states have been switched off, reduces to a particular case of Eq.(\ref{Eq.6.1}), of the presnt paper, provided that the substrate atom remains fixed, i.e 
$\varepsilon_m$,$V_{k m}$ (or $V_{m k}$) and, consequently, $\textsc{T}^{m}_{k k^{\prime}}$ are approximately time-independent. 
Then by the same methods employed in Sec.4 one can obtain Eq.(\ref{Eq.10.4}) in the $X\rightarrow 0$ limit. 
In Refs.\cite{Teillet,Makhmetov} the two state model
is used to explain the modification of singlet excited states $2^{1}S$ and $2^{1}P$ of $He^{\ast }$ interacting with an $Al$ surface. 
The problem is approached through the introduction of an effective $2 \times 2$, non Hermitian Hamiltonian \cite{Devdariani}, in the basis of the two, strongly
coupled, discrete states, in which the non Hermitian terms take into account the interaction with the continuum.

In this paper, we have integrated the reference Hamiltonian of two weakly coupled quasi-stationary states directly under the slowness hypothesis and we have proposed a manageable expression for the ionization probability that
can be easily compared with the well known result from the basic theory.
Thus we have estimated the influence of the new mechanism providing plausible arguments that substrate excitations produced by the primary ion bombardment during sputtering can, in some cases, alter the final charge-state of emitted particles.

\newpage

\section*{Figure Captions}

\begin{itemize}
\item  Fig.1:
Valence state of a sputtered atom from a metal surface, 
with time dependent energy $\varepsilon _{a}\left( t \right)$ and lorentzian broadening $\Delta _{a}\left( t \right)$, caused by interaction with an infinitely broad conduction band, with spectrum $\{ \varepsilon_k \}$, characterized by a Fermi level $\varepsilon_f$ and a work function $\phi$. 
Surface perturbations are modeled by the introduction of  a quasi molecular orbital, of energy  
$\varepsilon _{m}\left( t\right)$ and broadening 
$\Delta _{m}\left( t \right)$, transiently created
during the collision.

\item  Fig.2:
Electronic excitations generated by (Fig.2a) the Anderson-Newns resonant potential,(Fig.2b) the hoping interaction with the promoted MO and (Fig.2c) scattering from the screened cores of the transient QM.
(Fig.2d) first order correlation amplitude between the localized states 
$\left| a\right\rangle $ and $\left|m\right\rangle $.

\item  Fig.3: Collision cascade generated by a primary incident atom within the first layers of the surface region of a metal substrate. 
The number of secondary emitted particles is mainly determined by binary collisions with substrate atoms that have received a sudden impulse in the process and form Quasi Molecular systems with emitted atoms. 
The nature of these systems is transitory because the recoil velocity of substrate atoms moving within the metal is larger than the emission velocity.

\item  Fig.4: 
Schematic representation of the effective one-electron
potential (black line) of the first quantization Hamiltonian (\ref{Eq.oeham}) as a function of the normal position coordinate to the surface plane. 
Three main
contributions can be distinguished taking into account the effect of surface
barrier $v_{s}$, the sputtered atom $v_{a}$, and the quasi-molecule $v_{m}$.
Note that as the interatomic separation in the QM increases, the state  $\left| m\right\rangle$ approaches the continuum losing localization and the interacting terms $V_{km}$ and 
$V_{kk^{\prime }}$ become negligible.

\item  Fig.5: Comparison of Eqs.(\ref{Eq.10.5}), black line, and (\ref{Eq.61BNH}), gray line, as functions of the energy separation 
$\varepsilon _{af}=
\varepsilon_{a}-\varepsilon _{f}$ 
for a typical alkali sputtering experiment 
at $E=5$ eV of emission energy and $\mathrm{v}_b = 7.8 10^6$ cm/s of binding velocity.
The average outward velocity has been evaluated from Eq.(\ref{Eq.avvel}), the interaction length $\gamma ^{-1}$ set to $0.5$ \textrm{\AA} ,  the product $k_f X$ to $5$ and the
characteristic energy 
$\varepsilon _{1}$ to $10 \varepsilon_{0}.$

\item  Fig.6: 
Experimental data of Ref.\cite{Sroubek2} fitted, 
in the low velocity range $\mathrm{v}_{f} \preceq 6 \times 10^{6}$ cm/s, with a linear dependence of  $\mathcal{R}_{+}$ on $\mathrm{v}$ (black line). 
The binding velocity $\mathrm{v}_{b}$ has been used as a fitting parameter. 
Its result, in agreement with expected values, has been inserted into Eq.(\ref{Eq.34BN}) and used to fit data in the range 
$\mathrm{v}_{f} \succeq 9 \times 10^{6}$ cm/s (gray line).

\item  Fig.7: $\mathcal{R}_+$-dependence on temperatures in the case of dominant charge-exchange processes through the MO, for different values of the inverse of the average life-time of the QM, $\varepsilon_1$.
In each figure the black line has been obtained by numerically integrating Eq.(\ref{Important}) and inserting the result into Eq.(\ref{Eq.ionprobNa}) and 
the gray line represent the low temperature behavior of Eq.(\ref{Eq.10.4}).
$\varepsilon_0$ has been fixed for an emission energy of $1$ eV, with the $\mathrm{v}_b$ of Fig.5 and 
$\varepsilon_{af}$ to the literally value\cite{Sroubek2} $-0.4$ eV.
Despite a weak variation with increasing temperature, $\mathcal{R}_{+}$ is strongly sensitive to $\varepsilon _{1}$ and the range of validity of the analytical solution (\ref{Eq.10.4}) increases with increasing $\varepsilon_1$. 

\item  Fig.8: Comparison of different temperature behaviors of the ionization probability in the two limiting cases discusses in the article: 
The gray line corresponds to the numerical solution for $\mathcal{R}_+$, in absence of surface interactions, see  Eq.(\ref{Eq.61BNH}), and the black lines have been obtained as in Fig.7.
Parameters of Fig.7 have been used.
Note that the Anderson-Newns hopping mechanism seems to prevail at high
temperatures when $T\succeq 600$ K.
\end{itemize}
\pagebreak
\section*{figures}
\begin{figure}[htb]
     \begin{center}
     \epsfig{file=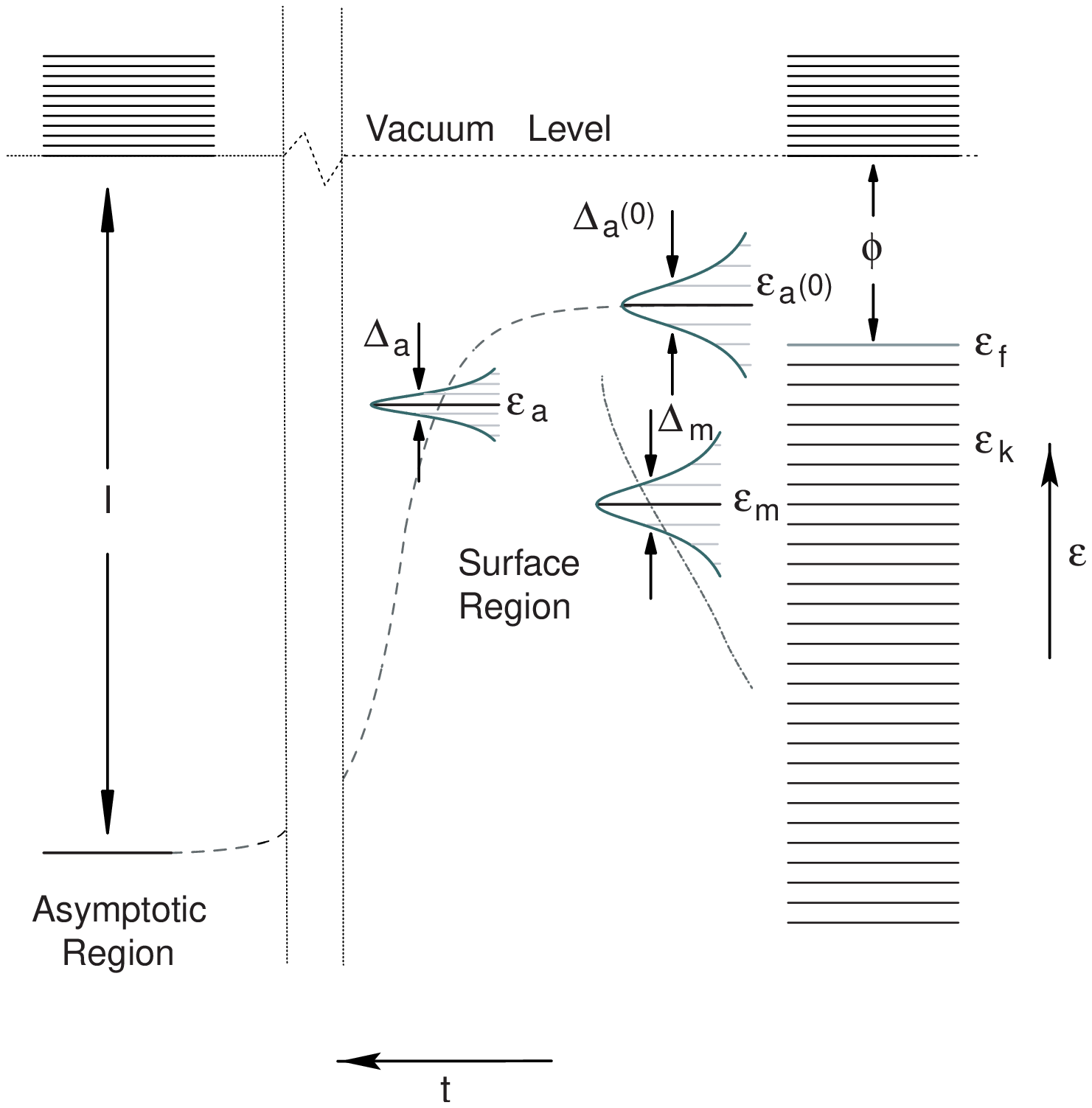,width=1.15\textwidth}
     \end{center}
     \caption{}
     \end{figure}
\pagebreak
\begin{figure}[htb]
     \begin{center}
     \epsfig{file=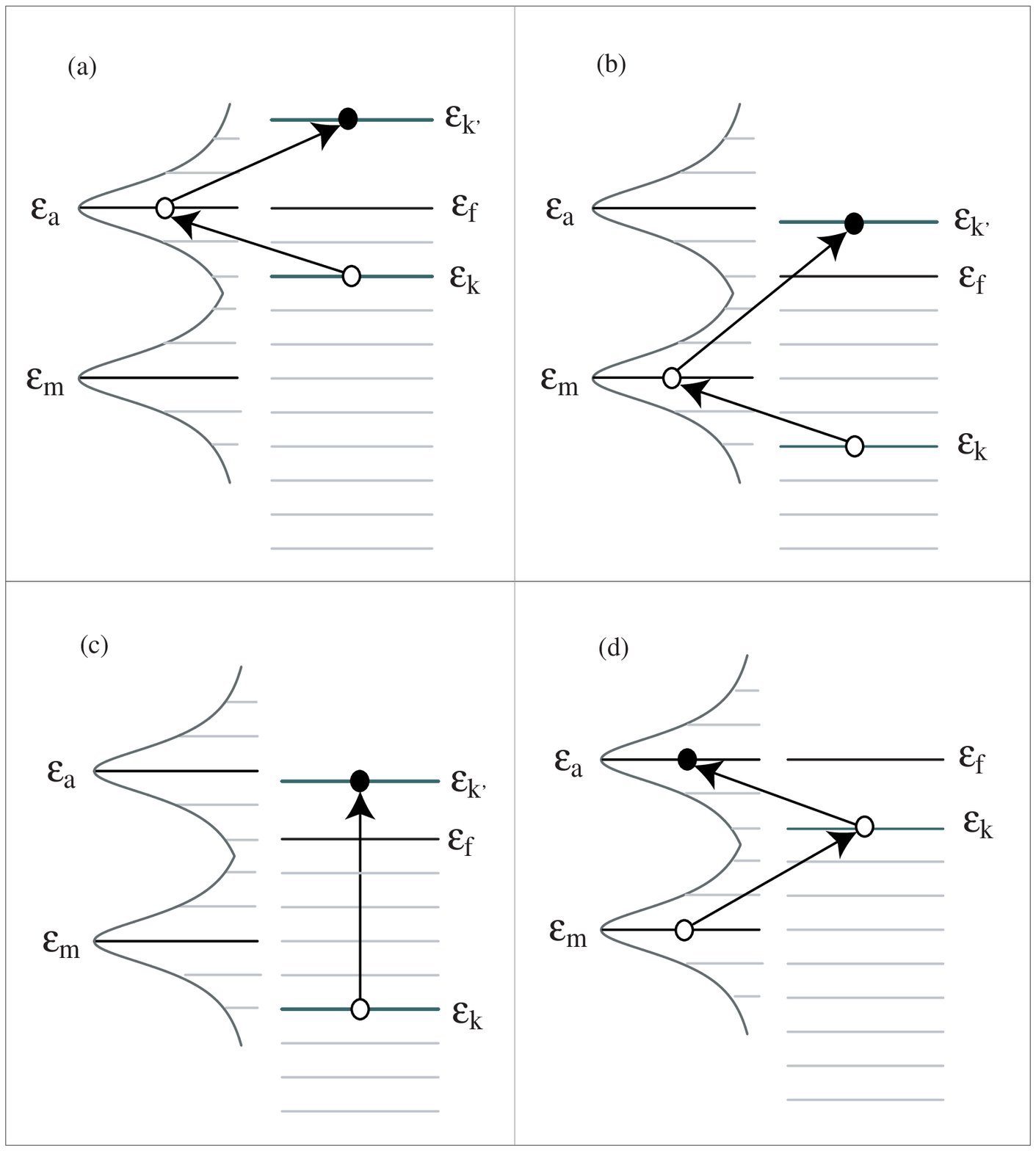,width=1.15\textwidth}
     \end{center}
     \caption{}
     \end{figure}
\pagebreak
\begin{figure}[htb]
     \begin{center}
     \epsfig{file=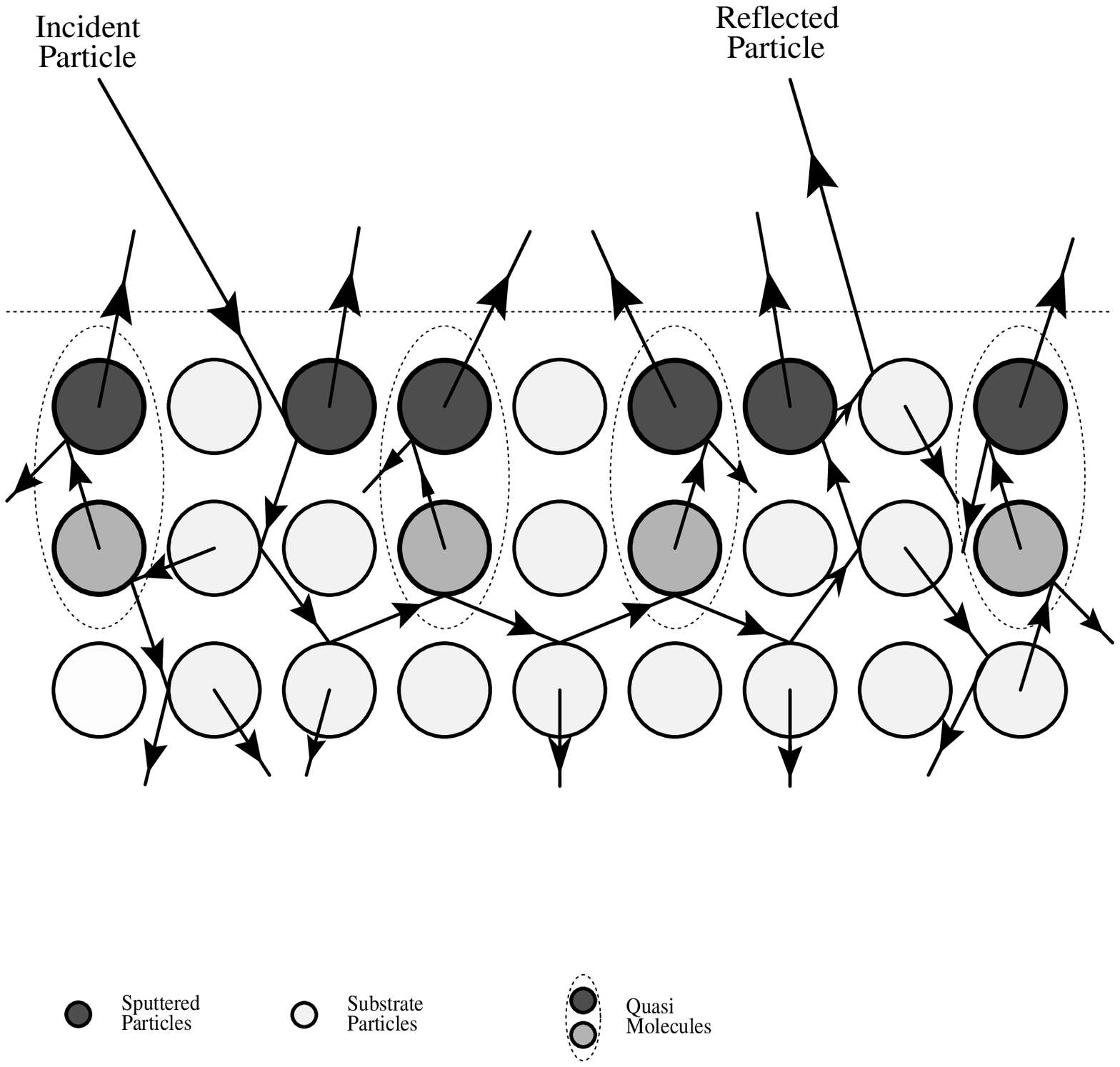,width=1.15\textwidth}
     \end{center}
     \caption{}
     \end{figure}
\pagebreak
\begin{figure}[htb]
     \begin{center}
     \epsfig{file=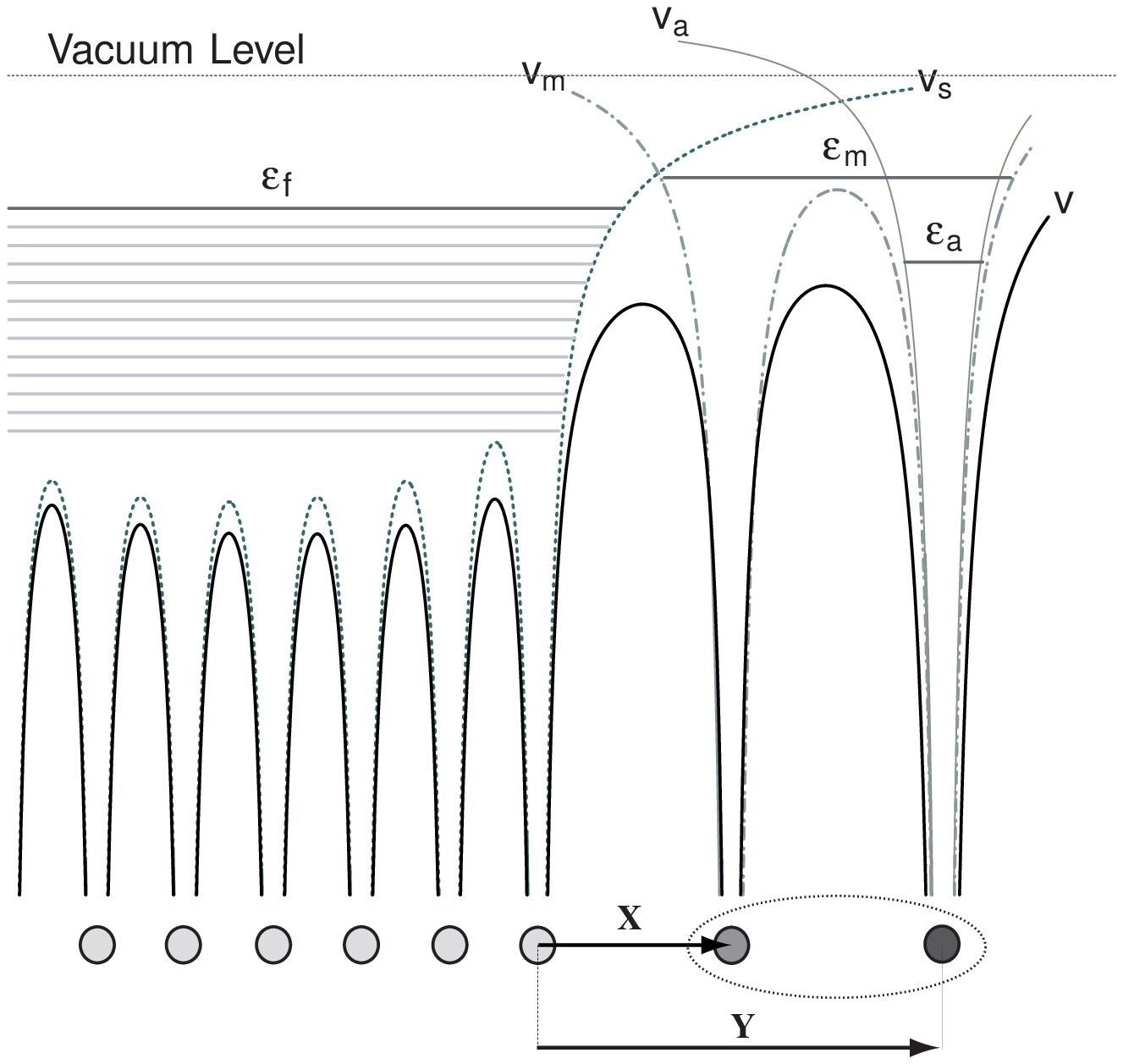,width=1.15\textwidth}
     \end{center}
     \caption{}
     \end{figure}
\pagebreak
\begin{figure}[htb]
     \begin{center}
     \epsfig{file=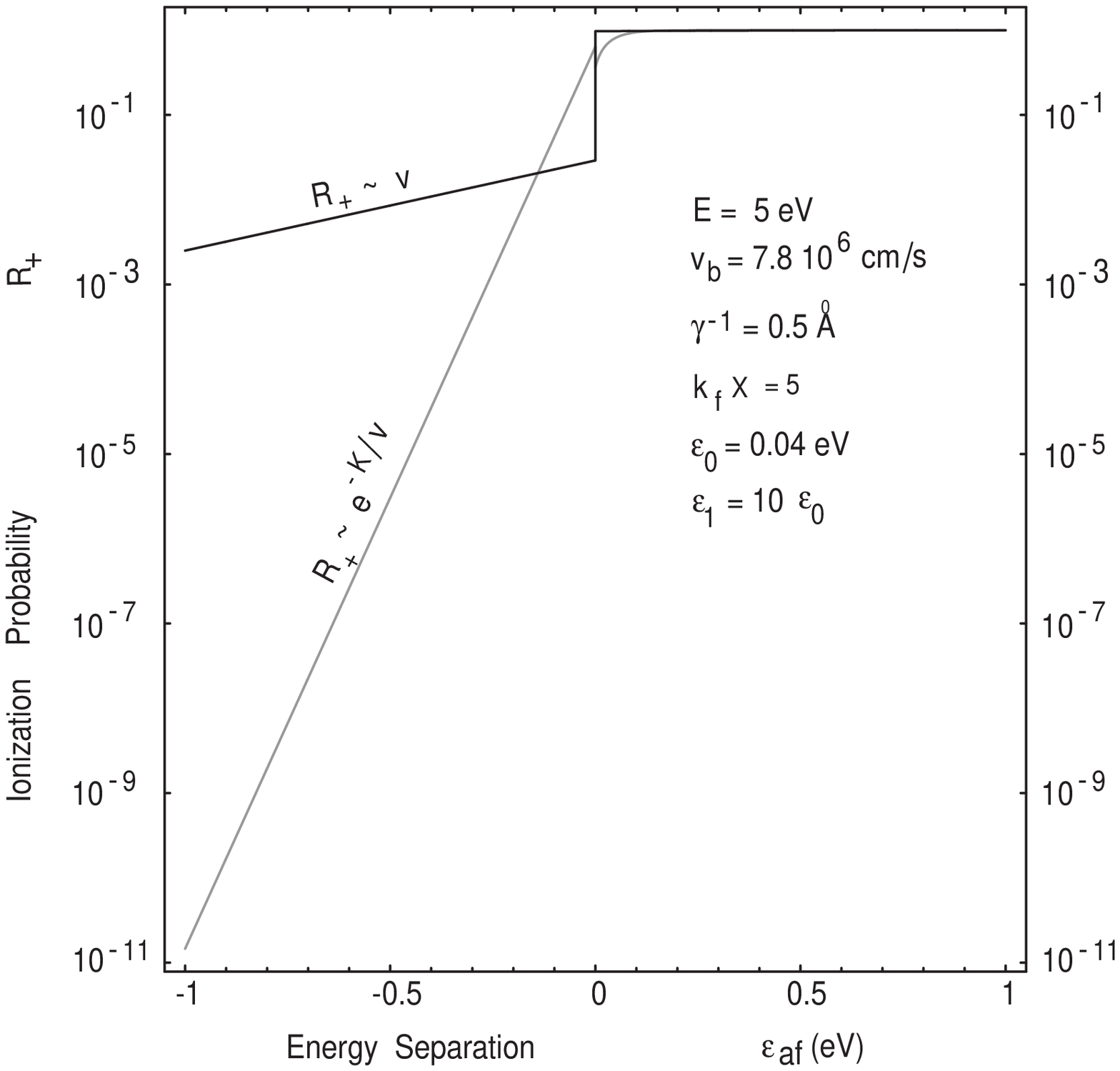,width=1.15\textwidth}
     \end{center}
     \caption{}
     \end{figure}
\pagebreak
\begin{figure}[htb]
     \begin{center}
     \epsfig{file=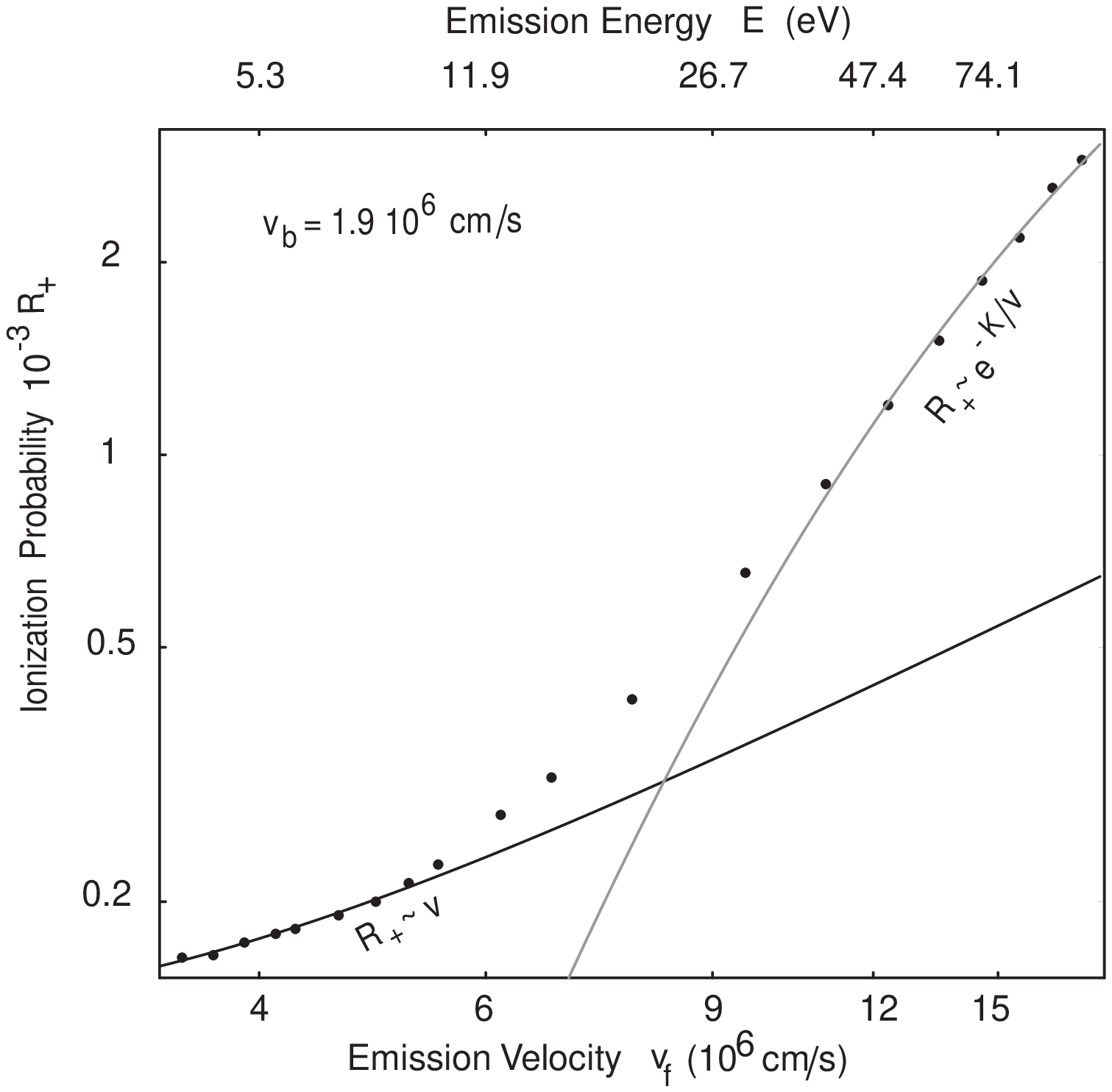,width=1.15\textwidth}
     \end{center}
     \caption{}
     \end{figure}
\pagebreak
\begin{figure}[htb]
     \begin{center}
     \epsfig{file=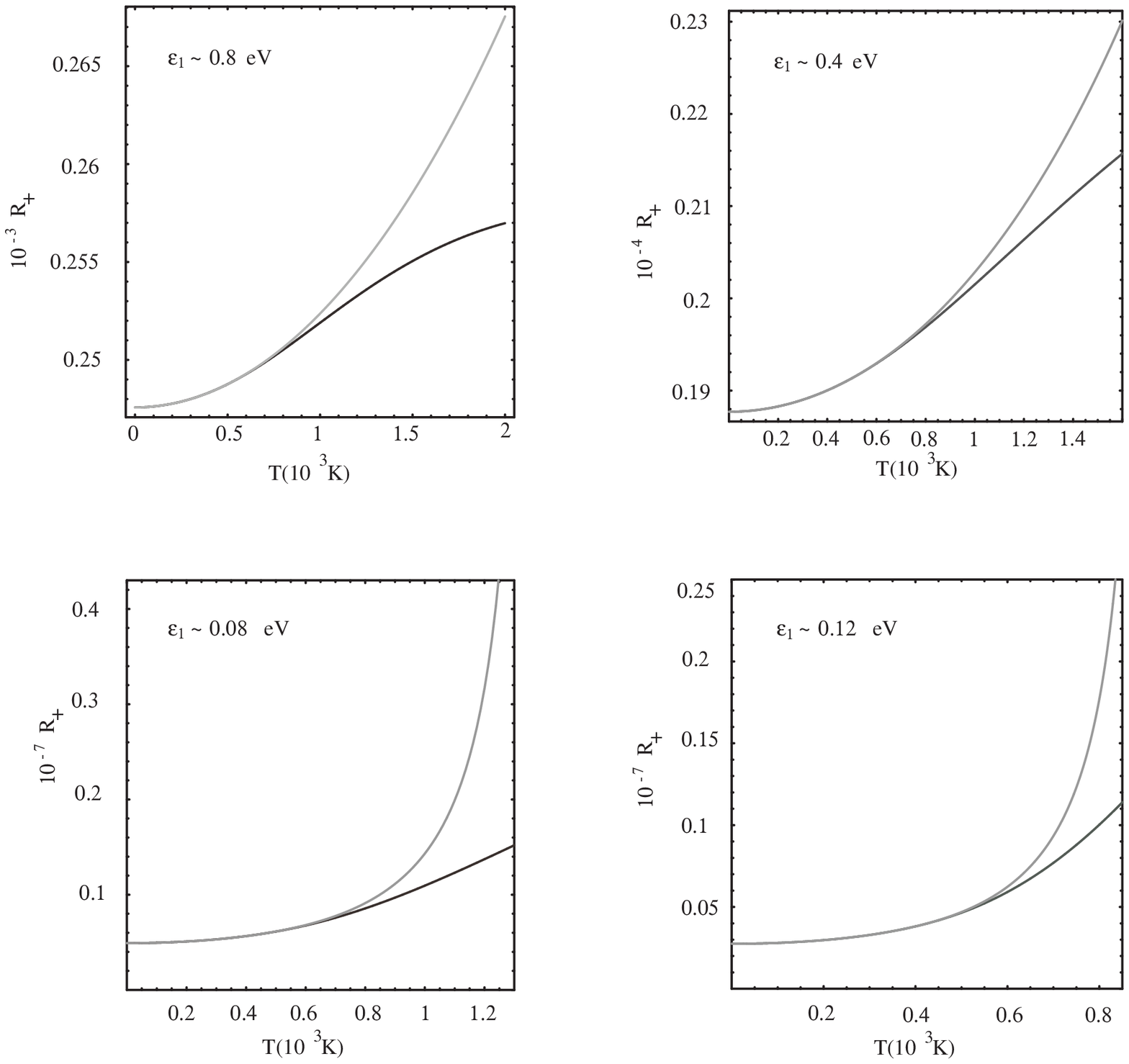,width=1.15\textwidth}
     \end{center}
     \caption{}
     \end{figure}
\pagebreak
\begin{figure}[htb]
     \begin{center}
     \epsfig{file=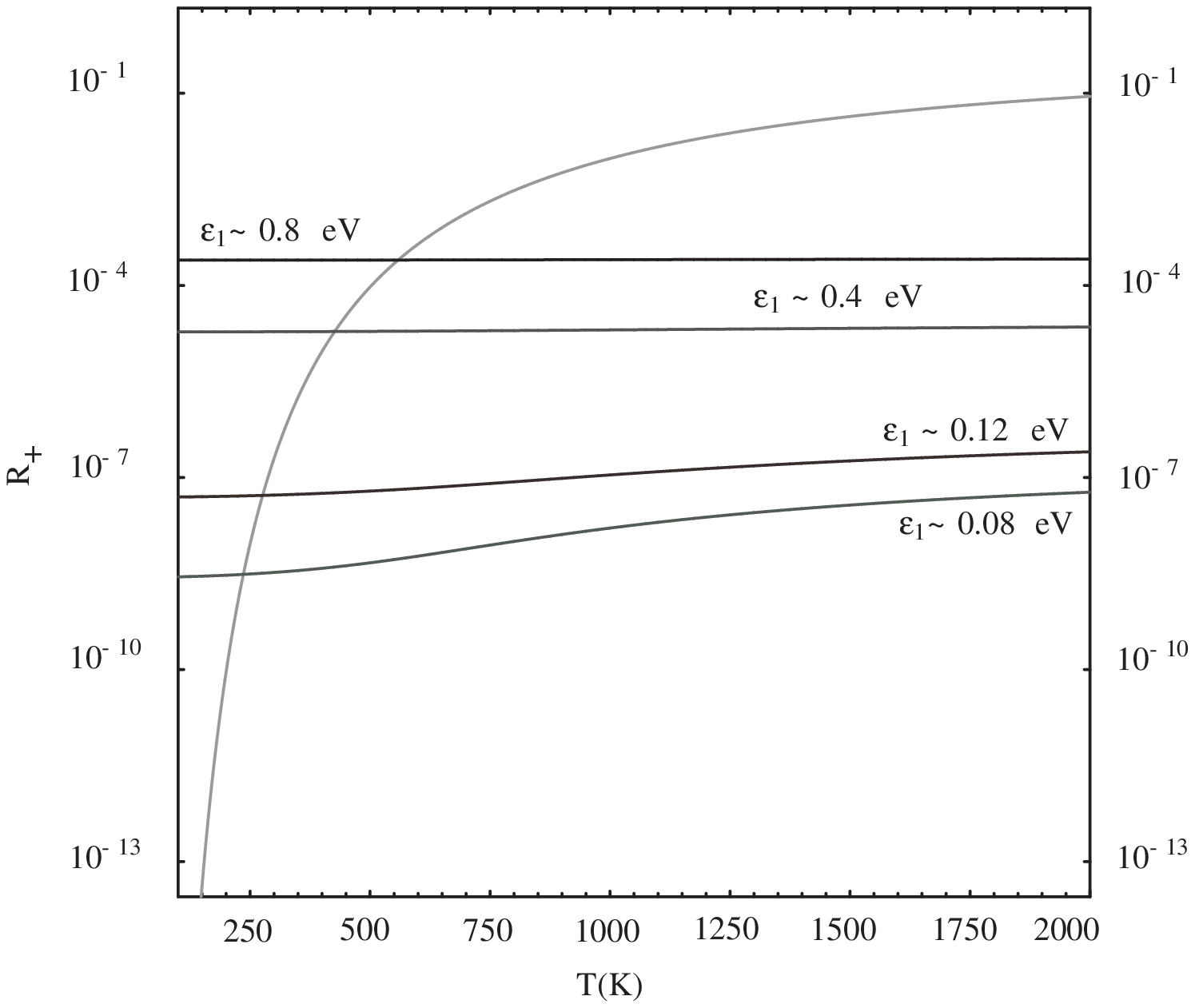,width=1.15\textwidth}
     \end{center}
     \caption{}
     \end{figure}
\end{document}